\documentclass[
  reprint,
  aps,
  pre,
  groupedaddress
]{revtex4-2}

\usepackage{hyperref}

\usepackage{amssymb}
\usepackage{amsmath}
\usepackage{bm}
\usepackage{enumitem}
\usepackage{graphicx}

\newcommand{\bv}[1]{\bm{#1}}

\newcommand{\mat}[1]{\underline{\bm{#1}}\mskip1.5mu}
\newcommand{\dt}{\delta t}

\newcommand{\PP}{ \mathbf{Pr} }
\newcommand{\EE}{\mathbb{E}}
\newcommand{\VV}{\mathbb{V}}
\newcommand{\RR}{\mathbb{R}}

\newcommand{\N}{\mathcal{N}}
\newcommand{\ra}{\rightarrow}
\newcommand{\pa}{\partial}
\newcommand{\tx}[1]{\textrm{#1}}
\newcommand{\dV}{\delta V}

\newcommand{\ind}{\!\perp \!\!\! \perp\!}
\newcommand{\iid}{i.i.d.}

\begin{document}


\title{Statistical field theory for dialectology}


\author{James Burridge}
\affiliation{School of Mathematics and Physics, University of Portsmouth, United Kingdom}


\date{\today}

\begin{abstract}
Is it possible to develop a `physics of language' which can explain the spatial, temporal and social patterns we see, and which can predict future change like we forecast the weather? Such a theory is likely to involve ideas from statistical physics. A substantial literature already applies these ideas to language. However, we lack a model which can match the spatial-temporal detail of historical changes at the level of individual linguistic features, and which offers a principled mechanism to predict the future. 
Here we present a statistical field theory for the evolution of linguistic variables which takes steps to fill this gap.  Linguistic variant frequencies are represented as a stochastic state field with spatial interaction and social conformity, coupled to a latent bias field with Onsager–Machlup action that reduces overfitting to data. We derive parameter inference procedures and demonstrate them using examples of large-scale dialect survey data from the twentieth century United States. The bias field has a characteristic half-life, which determines the horizon over which  linguistic change can be predicted. Inferred model parameters provide evidence for surface-tension-driven coarsening of dialect regions, with population-density gradients exerting systematic forces on interfaces. 
\end{abstract}


\maketitle

\section{Introduction}

\label{sec:intro}

Languages evolve constantly. This process may be described using \textit{linguistic variables} \cite{cha98,lab72} --- language features that take different forms, or \textit{variants}. Speakers within a population differ in their variant usage and these differences often exhibit strong spatial and social structure \cite{wie11,cha98,mil92,lab94,lab01,lab10}. When a group of speakers use the same variants of a distinctive set of variables, they are said to be using a \textit{dialect}. The fact that dialects are associated with geographical regions (or social groups) \cite{cha98} may be understood as a result of pattern formation within spatial (or social) linguistic variant distributions. Understanding how such patterns form, persist, and evolve remains a central problem in quantitative dialectology and the study of language change. 

The characterisation of dialect variation using descriptive statistical methods is well advanced \cite{gri11,jes24,doy14,kle24,wie15,tav16,rom22,fer23}. There are also \textit{non-spatial} mechanistic models with parameters that can be learned from data in a principled way using statistical inference \cite{mon23,mon23_2}. However, current \textit{spatial} models are not able to accurately describe, or to have their parameters inferred from, large scale modern linguistic survey datasets \cite{lee18,vau00}. As a result, there is a gap between descriptive spatial statistics and predictive dynamical models of language change. Here we bridge this gap by developing a statistical field-theoretic framework for modelling dialect evolution that is mechanistically grounded, inferable from data, and capable of describing observed patterns without overfitting.

\subsection{Language, genes and voters}

Analogies are often made between genetic and linguistic evolution \cite{gra03,bax05,rea10,bou12}. In the biological setting offspring inherit gene variants (alleles) from their parents, sometimes in mutated form \cite{eth11}. In the linguistic setting speakers inherit linguistic variants from their parents and others, and also mutate them. In both cases, each individual may be viewed as the descendent of a group (two parents in genetics, their speech community in linguistics), from which they inherited their variants.


A common modelling assumption in genetics and linguistics is that the probability  an individual will posses a given variant is equal to the relative frequency of that variant in the previous generation \cite{kim83,fis30,wri31}. Under this assumption, the evolution of relative frequencies may be described using the discrete time \textit{Wright-Fisher model} \cite{fis30,wri31}. The continuous time analogue of this model is \textit{Wright-Fisher diffusion} \cite{fel51,ewe04,eth11}. A closely related stochastic model, defined on a network, is the \textit{voter model} \cite{hol75,cli73,lig85}. Here, each node updates its variant by copying that of a randomly selected member of their network neighbourhood \cite{dor01}. Thus, the probability that a given node will adopt a given variant is equal to the relative frequency of that variant in its neighbourhood. The dynamics of sufficiently well connected groups of voters is approximated by the Wright Fisher diffusion model. If individuals form multiple densely connected clusters --- known as \textit{demes} in biology --- and are also able to migrate between them, we obtain a set of coupled Wright-Fisher diffusion processes known as the \textit{stepping stone model} \cite{Kim64, eth11}. We will refer to models in which variants are selected with probabilities equal to their relative frequencies as \textit{voter-type}. 

\subsection{Bias and conformity}

Historical language change datasets sometimes reveal sustained uni-directional shifts from one variant to another over time, known as `S-curves' \cite{lab94}. Such changes are implausible according to pure voter dynamics \cite{bly12}. A simple way to accommodate them is to allow a bias in the voter model which perturbs variant selection probabilities away from relative neighbourhood frequencies. The resulting  evolutionary model combines a stochastic component based on plausible assumptions about linguistic variant transmission, with the capability to model deterministic evolutionary forces. Such forces can arise from many sources, including social prestige associated with certain variants \cite{lab72,cha98}, the `learnability' of variants \cite{kir08,per11} or deliberate standardisation by authorities \cite{mil01}. 

\begin{figure}
 \centering
        \includegraphics[width=1\columnwidth]{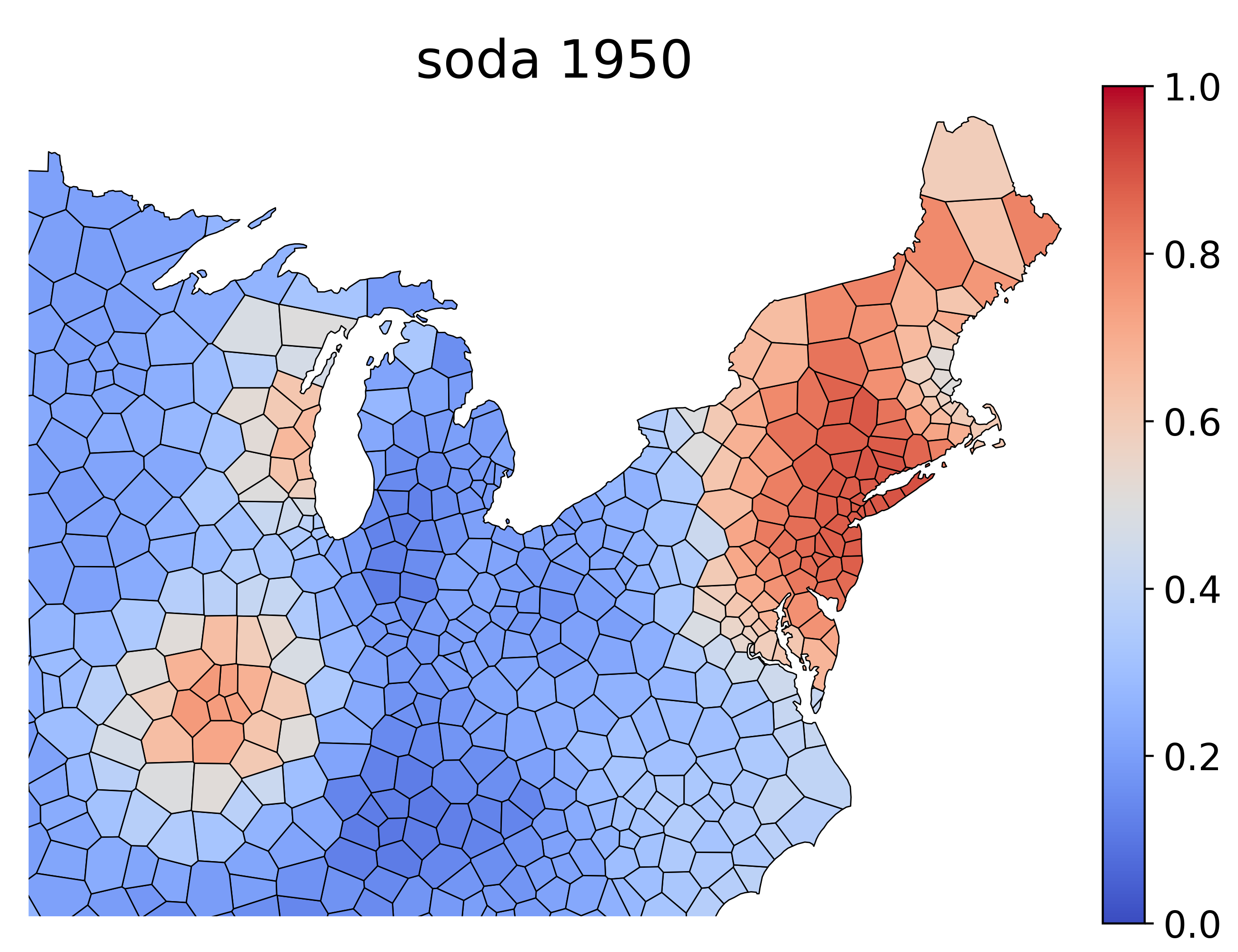}
 \caption{ Local logistic regression estimates of the fraction of speakers born in 1950 who use `soda' to describe a  carbonated beverage. Data from the Cambridge Online Survey of World Englishes \cite{vau00}. }
\label{fig:soda_zoom}
\end{figure}

As well as temporal S-curves, language datasets reveal  spatial regions in which one or more variants dominate \cite{lab06, cha98,gri19,bri09,lee19,mac22}. The boundaries of these domains, across which variant use may change quite sharply, are called \textit{isoglosses} \cite{cha98,wie15}. For example, Figure \ref{fig:soda_zoom} shows the geographical distribution of the term `soda' to describe a carbonated drink \cite{vau00}. Here we see isoglosses bounding Northeastern USA, St. Louis and Milwaukee. Although voter dynamics is capable of producing domains of this type, the conditions under which they emerge are  restrictive, requiring high noise and short range interactions, and the structure of their boundaries is complex and noise driven \cite{dor01}. An alternative mechanism is  social conformity. This may be modelled as a perturbation of pure voter dynamics which favours variants that are already popular. This (Ising-type) dynamics produces domain boundaries that behave like the surfaces of bubbles, being subject to surface-tension \cite{bra02, ham04, lip17, cas07}. 

\subsection{Dialect data and models}

The systematic study of dialect geography dates back at least to the 19th century \cite{wen76,ell89}. Older dialect surveys captured relatively few speakers. For example, the `Survey of English Dialects' \cite{ort78} a major project undertaken between 1950 and 1961 surveyed 986 respondents \cite{bur21} in 313 locations. Modern surveys, often carried out online, have generated much larger datasets. The English Dialect App \cite{lee18}, for example, collected over $5 \times 10^4$ responses, and the Cambridge Online Survey of World Englishes (the dataset we use) contains close to $10^5$ responses \cite{vau00}. 

Quantitative approaches to the study of dialect geography fall into two broad categories: \textit{statistical} and \textit{model based}.  Statistical studies infer geographical or social variations either in particular linguistic variables or in generalised metrics \cite{gri11,jes24,doy14,kle24,wie15,tav16,rom22,fer23}. Model based approaches aim to formulate the dynamics of variables that characterise language use  \cite{tak20,ise14,rad17,bur21,wil14,kau18, pro17,mon23,tru74,mur15, kau21,lou21,zan24, laz23}. These dynamical models  are sometimes exposed to data, and sometimes with exemplary rigour \cite{mon23,mon23_2}, but are not yet capable of matching data on long term spatial-temporal changes now available from dialect surveys. 

\subsection{Contribution of this paper}

There is a natural analogy between systems of interacting particles which can each exist in different physical states, and interacting people who can exist in different linguistic states. Models of language evolution, often inspired by statistical physics, have exploited this connection \cite{bax05,kau17,bar05,cas07,bar06}. Constructing population level models starting from simple assumptions about individual interactions is a promising approach because  behaviour in such systems at mesoscopic and macroscopic scales is often insensitive to many  of the details of microscopic interactions \cite{cas07,wil71}. Therefore, we don't need to get these details exactly right in order to make good predictions. While statistical physics inspired language models have been compared to data, a spatial evolutionary model which can resolve the observed mesoscopic patterns of historical change, with parameters that can be learned from data, is lacking. The construction of such a model is an important step toward understanding the \textit{physics of language}. Here we fill this gap by combining tools of statistical physics, which allow us to derive a mesoscopic spatial model of language dynamics from individual level interactions, with tools of statistical inference which allow us to learn the model's parameters \cite{has09,ras06,lin11}. The central methodological contributions of our work are
\begin{enumerate}[label=(\alph*)]
    \item A mesoscopic statistical field model for variant frequencies, respecting real population distributions, coupled to a latent bias field with constrained spatial flexibility, modelling historically contingent influences. 
    \item Use of the Onsager-Machlup  action \cite{ons53} for the bias field as a Bayesian prior, smoothing (regularising) inferred field trajectories to reduce overfitting and optimise its predictive power \cite{mur23,mur22}.
    \item Efficient inference methods allowing model parameters to be learned from large scale linguistic data-sets, revealing the relative importance of different microscopic mechanisms of linguistic change.
    \item The analytical formulation of interface dynamics, explaining observed patterns in terms of coarsening and population gradients. 
\end{enumerate}

The model and inference tools we develop represent a step towards a fully parameterisable general theory of language change.

\section{Model}

\begin{figure}
 \centering
        \includegraphics[width=1\columnwidth]{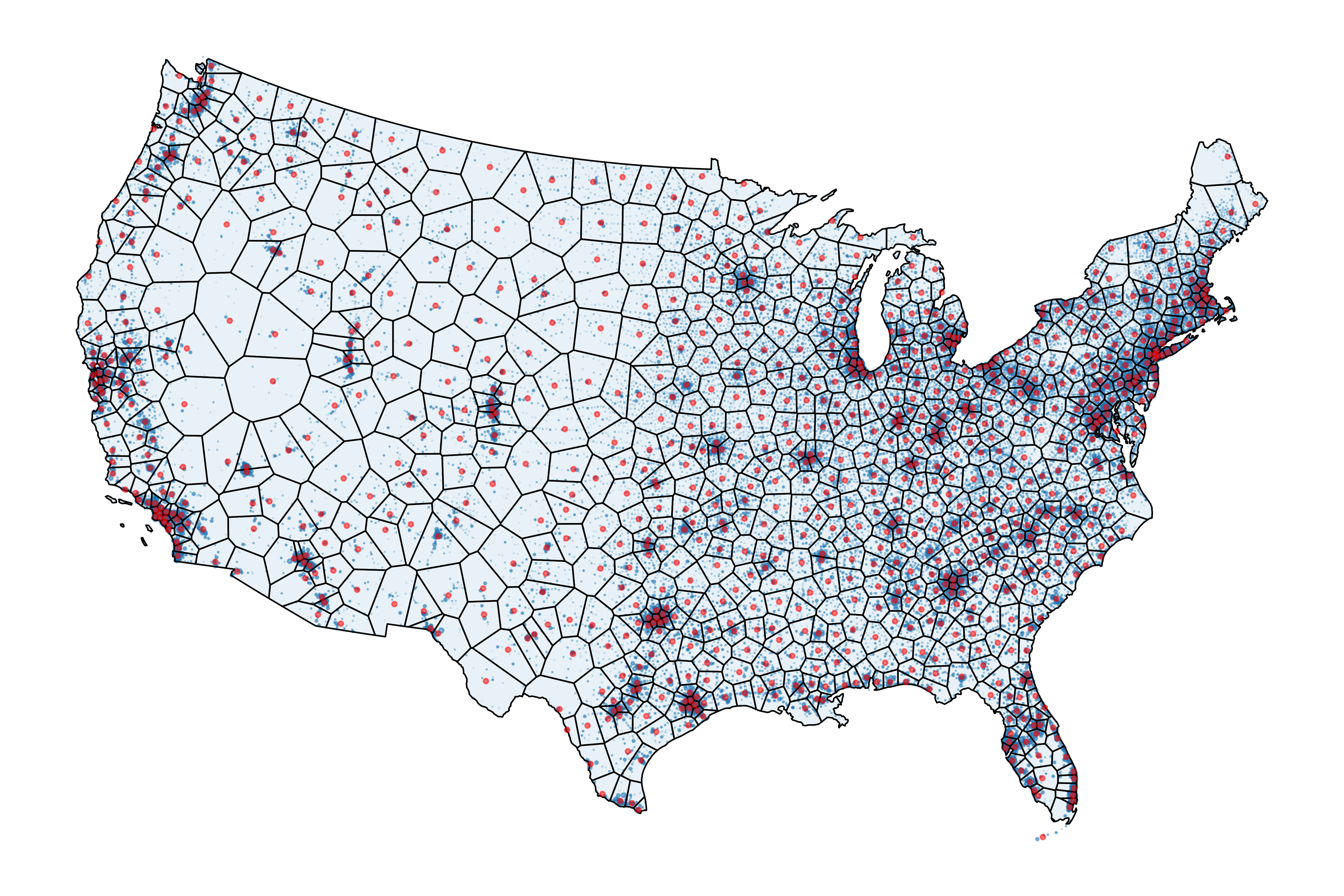}
 \caption{Voronoi tessellation of the USA obtained by k-means clustering citizens into $10^3$ demes based on their zip codes. Voronoi seeds are given by cluster centroids. We use the Universal Transverse Mercator EPSG:32615  coordinate system. Blue dots show zip codes with sizes proportional to the population at each code. Red dots show deme centroids. }
\label{fig:demes}
\end{figure}

Consider a geographical domain containing many residence sites, each  occupied by a single individual or `speaker'. For example a house might contain several sites, but a small flat only one.  We'll assume sites may be clustered into $n\geq 1$ groups of mutually `close' sites. We refer to each cluster, and the speakers within it, as a deme. We interpret `close' in a geographical sense, but our methodology  is in principle applicable when demes are defined using other proximity measures (e.g. social connectedness \cite{mil80}). The population in Figure \ref{fig:demes}  was obtained by applying the k-means algorithm to the set of all mainland USA zip codes, weighted by their populations. The mean population per deme is $329 \times 10^3$ with a standard deviation of $354 \times 10^3$. We will use this clustering throughout the paper,  denoting the centroid of the $i$th deme as $\bv{r}_i$.

\subsection{State field dynamics}

We consider a single linguistic variable with two variants $A$ and $a$ and let $X(\bv{r}_i,t)$ be the fraction of speakers in deme $i$  using variant $A$ at time $t$. Where spatial position is not needed explicitly within equations, we write $X(\bv{r}_i,t)=X_i(t)$. We write the vector of these fractions $\bv{X}(t)=(X_1(t), \ldots, X_n(t))$. We model the evolution of this `state field' using a family of coupled Wright-Fisher stochastic differential equations (SDEs)
\begin{align}
\nonumber
dX_i =& \Big( S_i(t)X_i(1-X_i) +  \beta (2X_i-1) X_i(1-X_i) \\
\nonumber
& + \nu(1-2X_i) +  \sum_{j=1}^n J_{ij} (X_j-X_i)\Big) dt \\
& + \sigma \sqrt{X_i(1-X_i)} dW_i   
\label{eqn:wf}    
\end{align}
where $W_1(t), \ldots, W_n(t)$ are independent standard Brownian motions \cite{oks03}, and $S_i(t)$ is the $i$th component of a stochastic bias field $\bv{S}(t) = (S_1(t), \ldots, S_n(t))$. We refer to $\bv{X}$ and $\bv{S}$ as fields because they vary over space and time. The interpretations of the terms in this model are given below. The family of SDEs (\ref{eqn:wf}) may be derived as a diffusion approximation to a speaker-level model within which each speaker uses either $A$ or $a$ at any given time, and speakers reselect their variants via a biased copying process based on the states of speakers in their own and other demes. This derivation is in appendix \ref{app:micro}. 

Models of similar form to (\ref{eqn:wf}) are well studied in statistical physics \cite{ham04} (when analysing critical phenomena using statistical field theory \cite{kar07}) and in genetics \cite{kim55}. Non-spatial Wright-Fisher models have been successfully deployed to model linguistic phenomena \cite{mon23,bax05}. The originality of our approach lies in our representation of space as a collection of linguistic demes, the introduction of a dynamical bias field (see section \ref{sec:bias}), and in out ability to infer model parameters from data. 

The terms in (\ref{eqn:wf}) have the following meanings. The \textbf{noise} term $\sigma \sqrt{X_i(t)(1-X_i(t))} dW_i$ is a diffusion approximation to the evolution of $X_i$ under the assumption that speakers in deme $i$ select variants $A$ and $a$ with probabilities $X_i$ and $1-X_i$ respectively. In isolation, this term approximates the dynamics of an unbiased voter model evolving in isolation within deme $i$. Within model (\ref{eqn:wf}), the noise term serves as approximation to the noise in a voter model perturbed by bias, locally conformist (Ising) dynamics, mutation and spatial interactions. The \textbf{interaction} term $\sum_{j=1}^n J_{ij} (X_j-X_i)$ accounts for the combined effect of deme-deme migration, and the influence on the variant selection processes of speakers in deme $i$ from speakers outside it. The parameter $J_{ij}$ is the sum of the migration rate into $j$ from $i$ and the `influence' of deme $j$ on deme $i$. We explain in appendix \ref{app:micro} why these two effects are not separately inferrable from linguistic data alone. The \textbf{bias} term $S_i(t) X_i(1-X_i)$ shifts the probability of selecting variant $A$ away from the current deme-average frequency $X_i$. Positive and negative values of $S_i(t)$ correspond to bias toward $A$, and $a$ respectively. The \textbf{conformity} term $\beta X_i(1-X_i)(2X_i-1)$, where $\beta \ge 0$, shifts the variant selection probabilities in favour of the currently more popular variant, mimicking social conformity. Larger $\beta$ produces stronger conformity and shifts the deme state faster toward universal use of one variant. The \textbf{mutation} term $\nu (1-2X_i)$ accounts for speakers randomly switching to the opposite variant at rate $\nu$ per unit time, without reference to the behaviour of others.

\subsection{Bias field dynamics}

\label{sec:bias}

Changing variables in model (\ref{eqn:wf}) to $X_i'=1-X_i$, yields a model for the dynamics of the proportion of variant $a$.  The conformity, mutation, interaction and noise terms are invariant under this change. As such we refer to them as \textit{neutral} processes --- they do not inherently favour or disfavour any particular variant \cite{bly10}. The bias term switches sign under the change, and is therefore \textit{non-neutral}. Our primary goal is to infer the  most probable form of the bias field from  linguistic data, leaving its interpretation to sociolinguists. Despite this flexibility of interpretation, there are still some reasonable and practical assumptions we can make about its dynamics. 

On a practical level, when performing inference we must be wary of \textit{over-fitting} \cite{bur26}. If fluctuations in the bias field are not suitably constrained then it will take over, spuriously, as the explanatory mechanism for historical changes. To avoid this we introduce hyperparameters which control spatial-temporal fluctuations.  Control of spatial fluctuations is achieved by defining the spatial field in terms of a lower dimensional `latent' field $\bv{\Psi}(t) \in \RR^m$ where $m \ll n$, and a \textit{lifting matrix} $\mat{A} \in \RR^{n\times m}$, as follows
\begin{align}
\label{eqn:S}
    \bv{S}(t) &= \mat{A} \bv{\Psi}(t) \\
\label{eqn:Psi}
    \dot{\Psi}_j(t) & = V_j(t) \\
    dV_j(t) &= -\frac{1}{\tau^2} (\Psi_j(t) + 2 \tau V_j(t))dt + \frac{2 \kappa}{\tau^{3/2}} dW_i. 
\label{eqn:V}
\end{align}
The field $\bv{\Psi}(t)$ may be interpreted as an embedding of the bias field in a lower dimensional space with the lifting matrix $\mat{A}$ acting to reconstruct $\bv{S}$ from this low dimensional representation. Representing a spatial field as a linear transformation of a lower dimensional latent field is known as the Linear Model of Coregionalization \cite{jou78}.  The lifting matrix is constructed so that fluctuations in the lifted field have a minimum length scale, set by a parameter $\eta$. The construction procedure is explained in section \ref{sec:lifting}. 

Equation (\ref{eqn:V}) may be recognised as the Langevin equation for the motion of a Brownian particle in a Harmonic potential well
$$
U(\Psi) = \frac{\Psi^2}{2 \tau^2} 
$$
subject to a viscous force $-2 V/\tau$, with the embedded field being the location of the particle. Modelling the bias field in this way imbues it with \textit{momentum}, meaning that the bias for or against variants will tend to increase or decrease in a sustained way over time. The harmonic potential well, which pulls the field back to zero, models the fact that trends of increasing or decreasing prestige associated with different linguistic variants tend eventually to end. The steeper the well the more short-lived trends will be. The noise parameter $\kappa$ may be interpreted as the propensity for trends to `get started'. 

A quantitative understanding of the effects of $\tau$ and $\kappa$ may be obtained as follows. The deterministic dynamics of $\Psi$ (setting $\kappa=0$) are
$$
\ddot{\Psi} + \frac{2}{\tau} \dot{\Psi} + \frac{1}{\tau^2} \Psi = 0
$$
with characteristic equation $\lambda^2+2\lambda \tau^{-2}+\tau^{-2}=0$. This has equal roots $\tau^{-1}$, corresponding to critical damping with timescale $\tau$. Solving (\ref{eqn:Psi}) and (\ref{eqn:V}) for the variance of $\Psi$ we find that in equilibrium $\VV(\Psi)=\kappa^2$. The parameters $\kappa$ and $\tau$ therefore control, respectively, the magnitude and time scale of bias fluctuations. Whereas the parameters of the state field dynamics, including the most likely form of the bias field, will be inferred directly from data, $\tau$ and $\kappa$ will be used as hyperparameters which specify our prior beliefs about the process which generated the field. They can also be used to control the complexity of the model (to `regularise' it \cite{has09}). In section \ref{sec:future} we will choose $\tau$ in order to maximise future predictive performance. 

It may be shown \cite{lin11} that the equilibrium solution to (\ref{eqn:Psi}) and (\ref{eqn:V}) is a zero mean Gaussian process with covariance 
$$
C(t,t')= \kappa^2 \left(1+\frac{|t-t'|}{\tau} \right) \exp\left(-\frac{|t-t'|}{\tau}\right).
$$
This is an example of a Mat\'{e}rn ($\nu=3/2$) covariance function.  An alternative interpretation of the bias field is as a functional parameter subject to a Gaussian process prior.

\subsection{Construction of lifting matrix}

\label{sec:lifting}

The role of the lifting matrix is to constrain the length scale, $\eta$, of fluctuations in the bias field. To construct $\mat{A}$ we define the Gram matrix \cite{mur22} of the radial basis function (Gaussian) kernel with length scale $\eta$
$$
W_{ij} = \exp\left(-\frac{\Vert \bv{r}_i-\bv{r}_j\Vert^2}{2 \eta^2} \right).
$$
Letting $D$ be a diagonal matrix with $D_{ii}=\sum_{j=1}^n W_{ij}$, then the `degree normalised' Gram matrix \cite{chu97} is defined
$$
\mat{\Sigma} = \mat{D}^{-1/2} \mat{W} \mat{D}^{-1/2}.
$$
This matrix is symmetric and positive definite with eigenvalues in the interval $[0,1]$ (see appendix \ref{app:LA}).  We can view it as the covariance matrix of a discrete, zero mean Gaussian random field defined on our demes with fluctuations of length scale $\eta$. Suppose we construct an orthonormal basis for $\RR^n$ such that the first $m <n$ basis vectors, $\bv{a}_1, \ldots, \bv{a}_m$ are sufficient to provide a near-faithful reconstruction, $\hat{\bv{S}}$, of any such random field, $\bv{S} \sim \mathcal{N}(\bv{0},\mat{\Sigma})$, as 
$$
\hat{\bv{S}} = \sum_{k=1}^m (\bv{a}_k^T\bv{S}) \bv{a}_k  =  \mat{A} \bv{B}.
$$
Here $\bv{B} \in \RR^m$ with $B_k = \bv{a}_k^T \bv{S}$ and $\mat{A} \in \RR^{n \times m}$ is the matrix whose columns are the basis vectors $\bv{a}_1, \ldots, \bv{a}_m$
\begin{equation}
\mat{A} = \begin{pmatrix}
| & | & & | \\
\bv{a}_1 & \bv{a}_2 & \ldots &\bv{a}_m \\
| & | & & |  
\end{pmatrix}.
\label{eqn:lift}    
\end{equation}
We have capitalised $\bv{B}$ to indicate that it is a  random (Gaussian) vector. The reconstruction $\hat{\bv{S}}$ will be faithful provided that the basis $\bv{a}_1, \ldots, \bv{a}_m$ is capable of describing arbitrary fluctuations of scale $\eta$. Conversely, using $\mat{A}$ as our lifting matrix will produce bias fields which fluctuate over the same scale. 

\begin{figure}
 \centering
        \includegraphics[width=1\columnwidth]{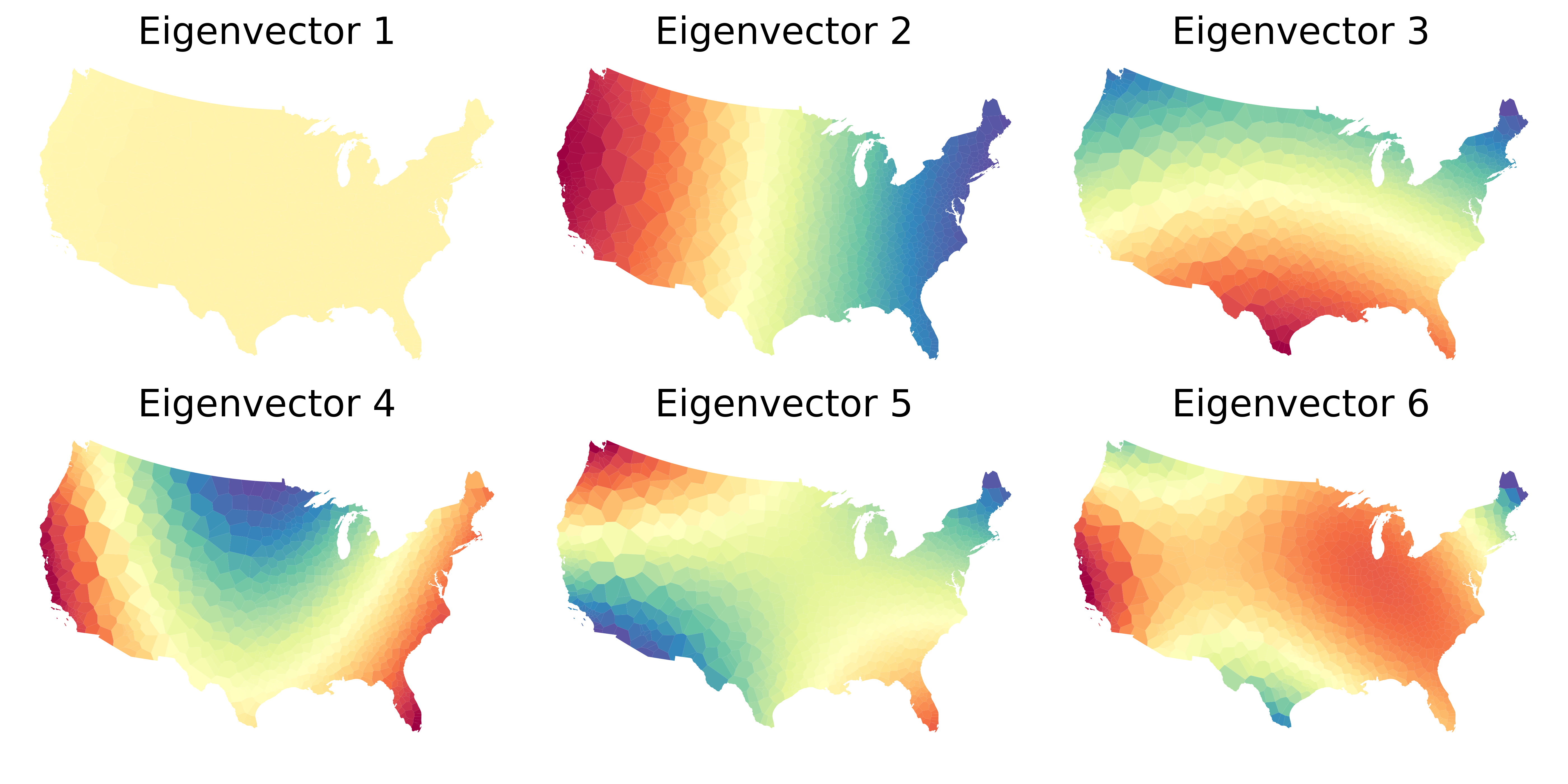}
 \caption{ The first six eigenvectors of $\mat{L}=\mat{I}-\mat{\Sigma}$ with $\eta=2000$km. In this case the variance retained is $R^2(6)=99.4\%$. }
\label{fig:recon}
\end{figure}

To construct the basis we define the Laplacian matrix $\mat{L}=\mat{I}-\mat{\Sigma}$, whose eigenvalues lie in the interval $[0,1]$. Since $\mat{L}$ is symmetric and positive semidefinite its eigenvectors $\bv{u}_1, \ldots, \bv{u}_n$ will be orthogonal with non-negative eigenvalues $\lambda_1, \ldots, \lambda_n$. We assume that the eigenvectors are normalised so  $\bv{u}_i^T\bv{u}_j=\delta_{ij}$ and ordered  so  $\lambda_{k+1}>\lambda_k$.  Suppose we use the eigenvectors of $\mat{L}$ as our basis, so $\bv{a}_k = \bv{u}_k$. Defining the reconstruction error 
$$
\Delta^2(m) = \EE(\Vert \bv{S}-\hat{\bv{S}}\Vert^2 ) = \sum_{k=m+1}^n \EE(B_k)^2,
$$
then the fraction of variation retained by our reconstruction is $R^2(m)=1-\Delta^2(m)/\Delta^2(0)$. We now observe that
\begin{align*}
    \EE(B_k^2) &= \EE\left( \sum_{i,j} V_i a_{ki} V_j a_{kj} \right) = \sum_{ij} a_{ki} \EE(V_i V_j) a_{kj} \\
    &= \bv{a}_k^T \mat{\Sigma} \bv{a}_k = \bv{a}_k^T (\mat{I}-\mat{L}) \bv{a}_k = 1 - \lambda_k.
\end{align*}
Therefore, the total variance retained by using the first $m$ eigenvectors of $\mat{L}$ as our basis is 
$$
R^2(m) = \frac{m-\sum_{k=1}^m \lambda_k}{n-\sum_{k=1}^n \lambda_k}.
$$
For given $\eta$ we construct our basis, and the lifting matrix, by selecting a minimum threshold, $R^2_{\ast}$, for the variance retained and letting 
$$
m = \min \left\{k | R^2(k)>R^2_{\ast}\right\}.
$$
For example, when $\eta=2000$km then $>99\%$ of the variance is retained using six eigenvectors (Figure \ref{fig:recon}).

Since the embedded field consists of $m$ independent Gaussian processes, then $\bv{S}=\mat{A} \bv{\Psi}$ will also be a Gaussian process. Provided $R^2(m)$ is close to one, then this lifting matrix provides a sufficient basis to represent bias fields which vary over a characteristic length scale $\eta$.  The spatial-temporal behaviour of the resulting bias field is determined by three hyperparameters $(\eta,\tau,\kappa)$ which control the scale of space and time fluctuations, and their overall magnitude. 

\section{Small noise approximation}

\label{sec:small_noise}

It is useful, when designing inference methods, to understand the behaviour of fluctuations in the state fields. Insights may be gained by considering their behaviour in the case of small noise \cite{gar09}. To condense notation we define the drift term in the state field model (\ref{eqn:wf}) as a function of the state vector and time
\begin{align*}
a_i(\bv{x},t) =& s_i(t)x_i(1-x_i) +  \beta (2x_i-1) x_i(1-x_i) \\
& + \nu(1-2x_i) +  \sum_{j=1}^n J_{ij} (x_j-x_i).   
\end{align*}
We then define $\bv{a}(\bv{x},t)=(a_1(\bv{x},t), \ldots, a_n(\bv{x},t))$ and the diagonal matrix $\mat{b}(\bv{x})$ with $b_{ij}(\bv{x},t)=\delta_{ij} \sqrt{x_i(1-x_i)}$. The state field model may then be written
\begin{equation}
d\bv{X} = \bv{a}(\bv{X},t) dt + \sigma \mat{b}(\bv{X}) d\bv{W}
\label{eqn:wfvec}
\end{equation}
where $\bv{W}(t)=(W_1(1), \ldots, W_n(t))$ is a vector of independent standard Brownian motions. Treating $\sigma$ as a small parameter, we write $\bv{X}=\bv{X}_0 + \sigma \bv{X}_1$ and substitute into (\ref{eqn:wfvec}), yielding
\begin{align*}
d\bv{X}_0 +  \sigma d\bv{X}_1 =& \left( \bv{a}(\bv{X}_0,t) +  \sigma D \bv{a}(\bv{X}_0,t) \bv{X}_1\right) dt \\
&+ \sigma \mat{b}(\bv{X}_0,t) d\bv{W} +O(\sigma^2)   
\end{align*}
where $D$ is the Jacobian operator, defined by
$$
\left[ D \bv{a}(\bv{x},t) \right]_{ij} = \frac{\pa a_i(\bv{x},t)}{\pa x_j}.
$$
Matching terms of of the same order in $\sigma$ we obtain
\begin{align}
\label{eqn:X0}
    d\bv{X}_0 &= \bv{a}(\bv{X}_0,t)dt \\
\label{eqn:X1}    
d\bv{X}_1 &= D \bv{a}(\bv{X}_0,t) \bv{X}_1 dt +  \mat{b}(\bv{X}_0,t) d\bv{W}.
\end{align}
From this we see that the state field may be understood as the sum of a deterministic field, $\bv{X}_0$ solving (\ref{eqn:X0}), plus a fluctuation field, $\sigma \bv{X}_1$.
Writing the diagonal terms of the Jacobian as $\left[ D \bv{a}(\bv{x},t) \right]_{ii} = -\theta_i(t)$, and defining $b_i(t)=\sqrt{X_{0i}(t)(1-X_{0i}(t))}$ then the SDE for the $i$th component of $\bv{X}_1$ is
$$
dX_{1i} = -\theta_i(t) X_{1i} dt + \sum_j J_{ij} X_{1j} dt + b_i(t) dW_i.
$$
Therefore in the small noise regime, fluctuations obey a set of linearly coupled Ornstein-Uhlenbeck process (\ref{eqn:X1}) with time dependent mean reversions $\theta_i(t)$ and volatilities $b_i(t)$. Due to the linearity of the processes, at this approximation order the state field is a Gaussian process with fluctuations scaling in proportion to $\sigma$ and spatial correlations determined by the range and strength of the interaction matrix $\mat{J}$. If the expected initial fluctuation is zero then it will remain zero for all $t$. Under these conditions the expectation of $\bv{X}(t)$ is equal to the deterministic field. This approximation becomes less effective for larger noise values, where non-linearities in the drift cause the expected and deterministic fields to differ more substantially. 

\section{Onsager-Machlup action for bias field}

\label{sec:OM_bias}

We will infer the bias field by searching for its most probable form given historical data. That is, we compute its maximum a posteriori (MAP) form. We first express the joint probability distribution of the field at a discrete mesh of times. In the continuum limit this yields the Onsager-Machlup action \cite{ons53} which will serve as a log-prior when computing MAP estimates.

Our starting point is an It\^{o} discretisation of the bias field dynamics over the time mesh $\{0,\Delta, 2 \Delta, \ldots, M \Delta\}$, written here for a single pair of fields $(\Psi,V)$,
\begin{align}
    \Psi_{n+1} -\Psi_n &= V_n \Delta \\
    V_{n+1}-V_n &= -\frac{1}{\tau^2}\left( \Psi_n + 2 \tau V_n \right) \Delta + \frac{2\kappa}{\tau^{3/2}} \sqrt{\Delta} Z_n
\end{align}
where $\Psi_n=\Psi(n\Delta)$, $V_n=V(n \Delta)$ and $Z_n \sim \N(0,1)$ is a standard normal noise variable with $Z_i \ind Z_j$ for all $i \neq j$. 
Eliminating the velocity field we obtain the following relationship between $Z_n$ and the bias field
\begin{equation}
    Z_n = \frac{\Delta^{1/2} \tau^{3/2}}{2 \kappa} \left( \ddot{\Psi}^{(\Delta)}_n + \frac{2}{\tau} \dot{\Psi}^{(\Delta)}_n + \frac{1}{\tau^2} \Psi_n\right)    
    \label{eqn:Z}
\end{equation}
where $\dot{\Psi}^{(\Delta)}_n$ and $\ddot{\Psi}^{(\Delta)}_n$ denote, respectively, the first and second discrete forward time derivatives of $\Psi$. Since the noise variables are \iid\ standard normal then the probability density of the vector $\bv{Z}=(Z_0, \ldots, Z_{M-2})$ is
$$
f_{\bv{Z}}(\bv{z}) = \frac{1}{(2\pi)^{\frac{M-1}{2}}} \exp \left(-\frac{1}{2} \sum_{n=1}^{M-2} z_n^2\right).
$$
We now view (\ref{eqn:Z}) as a change of random variables from $\bv{\Psi}=(\Psi_2, \ldots, \Psi_M)$ to $\bv{Z}$ with $\Psi_0, \Psi_1$ given by the initial conditions on the field and its first derivative.  The Jacobian matrix, $\mat{\mathcal{J}}$, of the transformation is causal (lower-triangular) with diagonal elements $\pa z_n/\pa \Psi_{n+2}$ independent of the field values. Therefore $|\mat{\mathcal{J}}|$ is a constant function of the parameters, independent of the fields. The density of the field is therefore
\begin{align}
\nonumber
f_{\bv{\Psi}}(\bv{\psi}) &= \\
& \frac{1}{\mathcal{Z}} \exp \left(-\frac{\Delta \tau^3}{8 \kappa^2} \sum_{n=0}^{M-2} \left(\ddot{\psi}^{(\Delta)}_n + \frac{2}{\tau} \dot{\psi}^{(\Delta)}_n + \frac{1}{\tau^2} \psi_n\right)^2\right),
\label{eqn:fPsi}
\end{align}
where $\mathcal{Z}$ is a normalising constant. The negative of the exponent of this density, in the limit of small mesh spacing with $T=M \Delta$, is the Onsager Machlup (OM) action functional
\begin{equation}
    S_{\tx{OM}}[\psi] = \frac{\tau^3}{8 \kappa^2} \int_0^T \left(\ddot{\psi}(t) + \frac{2}{\tau} \dot{\psi}(t) + \frac{1}{\tau^2} \psi(t)\right)^2 dt.
\label{eqn:SOM}
\end{equation}
Notice that the OM action involves the second time derivative of $\psi$, whereas only the first time derivative is defined along realised paths of the bias field. To see this, observe that $V(t)$ is an It\^{o} process, so not differentiable. In taking the limit $\Delta \ra 0$ of the density (\ref{eqn:fPsi}) we do not obtain a true density over paths (no such density exists \cite{das13}). Rather, the OM functional gives the probability that a random path $\Psi(t)$ will lie within a narrow `tube' of width $\epsilon$ around a smooth curve $\psi(t)$. To make this statement precise, let $w$ be a function defined on $[0,T]$, and define 
$$
\Vert w \Vert_T = \max_{t \in [0,T]} \vert w(t) \vert.
$$
According to this definition, if $f$ and $g$ are two functions defined on $[0,T]$  and we let $\epsilon = \Vert g-f\Vert_T$ then the constant width tube defined by the curves $f(t) \pm \epsilon$ for $t \in [0,T]$ is the narrowest such tube which also contains $g$. We will define the \textit{(epsilon) tube probability} of a smooth path to be the probability that a realised (rough) path will lie within a tube of radius $\epsilon$ around it. It may be shown (see \cite{ike81}) that for small $\epsilon$, there is a function $c(\epsilon,T)>0$, depending on $\epsilon$ and $T$ such that 
\begin{equation}
\PP\left(\Vert\Psi(t)-\psi(t)\Vert_T <\epsilon\right) \sim c(\epsilon,T) \exp\left(-S_{\tx{OM}}[\psi]\right)  
\label{eqn:POM}
\end{equation}
as $\epsilon \ra 0$. It follows that the ratio of the probabilities of a path $\Psi$ lying within a tube of width $\epsilon$ around smooth curves $\psi_1$ and $\psi_2$ is
$$
\frac{\PP\left(\Vert\Psi(t)-\psi_1(t)\Vert_T <\epsilon\right) }{\PP\left(\Vert\Psi(t)-\psi_2(t)\Vert_T <\epsilon\right) } = \exp\left(S_{\tx{OM}}[\psi_2]-S_{\tx{OM}}[\psi_1]\right).
$$
Using  (\ref{eqn:POM}) as a prior on the bias field and maximising the posterior over smooth paths results in a smooth path which has maximum posterior tube probability, asymptotically as $\epsilon \ra 0$.

\section{Inference from linguistic data}

\label{sec:inf}

Our goal is to infer the parameters of our model from data, allowing us to explore model behaviour analytically and through simulation, and make predictions. Ideally we would have observations of many speakers in each deme at many times. In reality, such longitudinal data is difficult to collect, requiring repeated surveys of the same community over years. 

\subsection{Apparent time}

In the absence of longitudinal data we can use the \textit{apparent time} principle \cite{cha98}, which assumes that the linguistic state of post-adolescent speakers can be used to approximate the state of the language community to which they belonged during their adolescence. The principle is based on the observation that once speakers pass early adulthood their linguistic systems tend to stabilize. Language change in the community as a whole is therefore  driven by younger speakers, whose rates of copying and mutation are much higher. Although the apparent time principle has limitations, notably that adult language is not fully stable, comparison of real and apparent time observations support its assumptions \cite{bai91,san05,lab94,san18}.

Our model assumes that all sites update at the same frequency. According to the apparent time principle, update rates will vary between sites, according to the age of the speakers they contain. Rather than explicitly incorporating these variations into our dynamics, we  interpret the update rate at each site as the average rate of the population as a whole. 

We make the definition of apparent time precise as follows. Suppose we have one or more linguistic surveys collected on a variety of dates. Suppose also that the birth dates of the respondents were recorded. The collection date of each response gives a \textit{real time} reading of the linguistic state of the community. The birth date of the respondent gives an \textit{apparent time} reading of the community they grew up in.  In what follows we use birth date as our time variable.


\subsection{Posterior parameter distribution}

For each binary linguistic variable, the dataset for deme $k$ consists of three vectors
\begin{align*}
    \bv{t}_k &= (t_{k1}, t_{k2}, \ldots, t_{km_k}) \\
    \bv{n}_k &= (n_{k1}, n_{k2}, \ldots, n_{km_k}) \\
    \bv{y}_k &= (y_{k1}, y_{k2}, \ldots, y_{km_k}),
\end{align*}
where $m_k$ is the number of different observation times for the deme, $n_{ki}$ is the number of respondents from that deme at time $t_{ki}$, and $y_{ki}$ is number of these respondents who report using variant $A$. We also write $\bv{x}_k=(x_{k1},x_{k2}, \ldots, x_{km_k})$ for the true fractions of $A$ speakers at the observation times. Viewing $\bv{x}_k$ and $\bv{y}_k$ as realisations of random vectors $\bv{X}_k$ and $\bv{Y}_k$, then the conditional probability mass function of $\bv{Y}_k$ given $\bv{X}_k$ is 
$$
f_{\bv{Y}_k|\bv{X}_k}(\bv{y}|\bv{x}) = \prod_{i=1}^{m_k} \binom{n_{ki}}{y_{ki}} x_{ki}^{y_{ki}}(1-x_{ki})^{n_{ki}-y_{ki}}. 
$$
For a given realisation, $\bv{s}$, of the bias field, the parameters of our model (\ref{eqn:wf}) may be written as a vector $\bv{\theta}=(\bv{s},\beta,v,\mat{J},\sigma)$. 
We view $\bv{\theta}$ as a realisation of the latent variable $\bv{\Theta}$, with prior distribution $f_{\bv{\Theta}}(\bv{\theta})$.

Letting $\bar{x}_k(t;\bv{\theta})$ be the zero noise solution to our model for deme $k$ then the state vector $X_k(t)$ is given by
$$
X_k(t) = \bar{x}_k(t;\bv{\theta}) + \delta X_k(t) 
$$
where $\delta X_k(t)$ is stochastic fluctuation process. These fluctuations are not  directly observable, but nevertheless we can define a vector giving their values at the observation times 
$$
\bv{\delta x}_k=(\delta x_{k1}, \ldots, \delta x_{km_k}),
$$ 
which we view as a realisation of the random vector $\bv{\delta X}_k$. As noted in section \ref{sec:small_noise}, these fluctuations are Gaussian with zero mean under the small noise approximation. For larger noise values they will bias away from zero as the expected state field departs from its zero noise form. When devising fast inference methods below, we will assume that the small noise approximation $\EE(\delta X_{ki}) \approx 0$ holds. 

Defining the concatenations, $\bv{\delta X}$ and $\bv{Y}$, of the fluctuations and observations in every deme, then the probability mass function of the observations given the fluctuations and the parameters is
\begin{align}
\nonumber
&f_{\bv{Y}|\bv{\delta X},\bv{\Theta}}(\bv{y}|\bv{\delta x},\bv{\theta}) =  \\
&\prod_{k=1}^n \prod_{i=1}^{m_k} \binom{n_{ki}}{y_{ki}} (\bar{x}_{ki}+\delta x_{ki})^{y_{ki}}(1-\bar{x}_{ki}-\delta x_{ki})^{n_{ki}-y_{ki}}.   
\end{align}
This is the (conditional) \textit{likelihood function} of the parameters. We will assume flat priors for all parameters other than the bias field, whose prior is given by the measure of the process defined by (\ref{eqn:S}), (\ref{eqn:Psi}) and (\ref{eqn:V}). The log-posterior is then
\begin{align}
\nonumber
&\log f_{\bv{\Theta},\bv{\delta \bv{X}}|\bv{Y}}(\bv{\theta},\bv{\delta x}|\bv{y})= \sum_{k=1}^n \sum_{i=1}^{m_k} \Big( y_{ki} \log (\bar{x}_{ki}+\delta x_{ki}) \\
\nonumber
& + (n_{ki}-y_{ki}) \log (1-\bar{x}_{ki}-\delta x_{ki} \Big) + \log f_{\bv{\delta X}|\bv{\Theta}}(\bv{\delta x}|\bv{\theta}) \\
&  - \sum_{i=1}^m S_{\tx{OM}}[\psi_i] + \tx{const.}
\end{align}
In principle we could obtain MAP estimates for both the parameters and the fluctuations by finding the values of $\bv{\theta}$ and $\bv{\delta x}$ which maximise this expression. However in practice, due the high dimensionality of the optimisation problem, and the lack of a closed form expression for the density of the fluctuations, $f_{\bv{\delta X}|\bv{\Theta}}$, this is an impractical calculation. We therefore proceed in two stages. First we estimate the `deterministic' parameter vector, $\bv{\theta}_0$, which excludes $\sigma$.  We then estimate $\sigma$ via a separate procedure, based on our estimate, $\hat{\bv{\theta}}_0$. To estimate $\bv{\theta}_0$ we maximise the `quazi-log-posterior' obtained by setting the fluctuations in the likelihood to zero 
\begin{align}
\nonumber
&\log f^{(Q)}_{\bv{\Theta}_0|\bv{Y}}(\bv{\theta}_0|\bv{y})= \\
\nonumber
&\sum_{k=1}^n \sum_{i=1}^{m_k} \Big( y_{ki} \log (\bar{x}_{ki}) + (n_{ki}-y_{ki}) \log (1-\bar{x}_{ki}\Big) \\
&  - \sum_{i=1}^m S_{\tx{OM}}[\psi_i] + \tx{const.}
\end{align}
Here we have approximated the log probability density of the fluctuations as a constant (with respect to $\bv{\theta}_0$) on the grounds that the sensitivity of their statistics to variations in the deterministic trajectory around its MAP form will be negligible. 

We now explain the conditions under which we expect maximising the quazi-log-posterior to provide accurate inferences. We test the consistency of our procedure  in section \ref{sec:USA}. We define the radius of a typical deme as
$$
R_{\tx{deme}} = \sqrt{\frac{A}{n \pi}}
$$
where $A$ is the total land area surveyed. In the case of mainland USA, $A=7.67\times 10^6 \tx{km}^2$, so with $n=10^3$ demes, $R_{\tx{deme}} \approx 50$km. Field observations and surveys  \cite{lab06,cha98,vau00} show that `uniform' regions in which variant frequencies are more or less constant are typically larger than $R_{\tx{deme}}$, and that the time scale of systematic shifts in variant usage is of the order of decades at least \cite{cha98} (see also Figures \ref{fig:soda} and \ref{fig:roly}). Within these larger stable regions and time scales we also expect to see small local fluctuations. Let $R_{\tx{uni}}$ be the typical length scale of uniform variant usage regions, and  $T_{\tx{sta}}$ be the time scale over which these uniform regions remain stable.  Writing our state field
$$
\bv{X} = \bar{\bv{x}} + \bv{\delta X}
$$
we identify long length scale variations with $\bar{\bv{x}}$ and short length scale variations with $\bv{\delta X}$. Let $\mu$ be the number of observations made per unit area per unit time in our dataset and suppose that
$$
T_{\tx{sta}} R_{\tx{uni}}^2 \gg \frac{1}{\mu}.
$$
In this case we will have in our dataset many observations generated by state fields which differ only due to fluctuations around a near-constant deterministic field $\bar{\bv{x}}$, which we describe as \textit{densely sampled}. Suppose also that our model is well-specified in the sense that there exists a parameter vector, $\bv{\theta}^\ast$ for which our observations can be considered to be a sample from  the model. That is
\begin{equation}
Y_{ki}|\delta X^\ast_{ki} \sim \tx{Binomial}(n_{ki},\bar{x}_{ki}^\ast + \delta X^\ast_{ki}),    
\label{eqn:Yki}
\end{equation}
where $\bar{x}_{ki}^\ast$ is the deterministic field under the model with $\bv{\theta} = \bv{\theta}^\ast$ and $\delta X^\ast_{ki}$ is a random fluctuation in the same model. We will refer to this as the `true' distribution of $Y_{ki}$. Under dense sampling there will be many observations generated using approximately the same deterministic field, but different realisations of the fluctuations. Writing $y_{ki}=\bar{x}^\ast_{ki}+\delta y_{ki}$ then, for example
$$
y_{ki} \log(\bar{x}_{ki}) = \bar{x}^\ast_{ki} \log(\bar{x}_{ki}) + \delta y_{ki} \log(\bar{x}_{ki}).
$$
The expected value of each $\delta y_{ki}$ will depend on the unobservable fluctuations. However, because there will be many similar `delta-$y$' terms in the quazi-log-likelihood corresponding to approximately the same deterministic pair $(\bar{x}^\ast_{ki}, \bar{x}_{ki})$, their sum  will be $\approx 0$. Therefore each observation $y_{ki}$ can be replaced with its expectation under the true model, yielding the following approximation of the quazi-log-posterior
\begin{align}
\nonumber
&\log f^{(Q)}_{\bv{\Theta}_0|\bv{Y}}(\bv{\theta}_0|\bv{y}) \approx \\
\nonumber
&\sum_{k=1}^n \sum_{i=1}^{m_k} n_{ki} \Big( \bar{x}^\ast_{ki} \log (\bar{x}_{ki}) + (1-\bar{x}^\ast_{ki}) \log (1-\bar{x}_{ki}\Big) \\
&  - \sum_{i=1}^m S_{\tx{OM}}[\psi_i] + \tx{const.}
\label{eqn:qlp}
\end{align}
The summation over state fields in (\ref{eqn:qlp}) may be recognised as a sample approximation to the cross entropy between two different fluctuation-free models of variant choice according to which the probabilities of selecting variant $A$ at time $t$ in the individual demes are given by $\bar{\bv{x}}^\ast(t)$ and $\bar{\bv{x}}(t)$. 

When maximising (\ref{eqn:qlp}) with respect to $\bv{\theta}$, the OM action constrains smoothness of the bias field. Provided $\bv{s}^\ast$ lies within the plausible region defined by the OM action (the OM hyperparameters are well chosen) then $\mat{\bv{\theta}}$ will minimise the cross extropy between the model and the true model, so $\hat{\bv{\theta}} \approx \bv{\theta}^\ast$. 

To summarise, provided our data is sufficiently densely sampled and our prior well chosen, then the maximum of the true log-posterior and the quazi-log-posterior will approximately coincide because the total contribution due to the unobservable fluctuations will be negligible.

\subsection{Estimating the noise parameter}

To estimate $\sigma$ we exploit the fact that fluctuations around the deterministic field cause the observation variable $Y_{ki}$  to be \textit{overdispersed} with respect to its zero noise form. To see this, consider a single space-time point with deterministic field $\bar{x}$, and consider the observation-fluctuation pair $(Y,\delta X)$, where $Y|\delta X \sim \tx{Binomial}(n,\bar{x}+\delta X)$. Assuming that $\EE(\delta X) =0$ then by the law of total variance 
\begin{align}
\nonumber
\VV(Y) &= \EE(\VV(Y|\delta X))- \VV(\EE(Y|\delta X)) \\
&= n \bar{x}(1-\bar{x}) + n(n-1) \VV(\delta X),   
\label{eqn:overd}
\end{align}
where $\VV(\delta X) \propto \sigma^2$. From this we see that the excess variance of our observations over its zero noise value is proportional to $\sigma^2$. Based on this observation we define the `excess variance' statistic
$$
T(\bv{Y}; \bv{\theta}_0) = \frac{\sum_{k,i} \left((y_{ki}-\bar{x}_{ki})^2 - n_{ki} \bar{x}_{ki}(1-\bar{x}_{ki})\right)}{\sum_{k,i} n_{ki}(n_{ki}-1)} 
$$
which can be computed using only the observations and the deterministic parameters. Given an estimate $\hat{\bv{\theta}}_0$ we write the value of this statistic on observed data as
$$
\hat{T} = T(\bv{y};\hat{\bv{\theta}}_0).
$$
We can also compute the expectation of $T$ over all possible realisations of the observations for given $\hat{\bv{\theta}}_0$ and $\sigma$ via Monte Carlo integration. We simulate the model many times, sampling the observations from each simulation using the same values of $n_{ki}$ as in the real data. This yields the expected statistic
$$
\bar{T}(\hat{\bv{\theta}}_0,\sigma) = \EE_{\sigma}(T(\bv{y};\hat{\bv{\theta}}_0))
$$
where $\EE_\sigma$ denotes expectation with respect to the model with noise parameter $\sigma$. We then estimate $\sigma$ as the solution to
$$
\bar{T}(\hat{\bv{\theta}}_0,\sigma) = \hat{T}.
$$
Note that estimating $\sigma$ in this way is most effective when  $n_{ki} \gg 1$. Demes with $n_{ki}=1$ are uninformative since in this case overdispersion is absent (see equation (\ref{eqn:overd}). This places constrains the maximum number of demes that can be used if we want to infer $\sigma$. 

\section{Simplified interaction model}

In its most general form the interaction matrix $\mat{J}$ contains $n(n-1)$ free parameters. Even with substantial quantities of observational data, allowing these parameters to vary independently will lead to serious overfitting. In the current paper we consider a simplified interaction model with a single learnable parameter, $J$. Letting $N_i$ be the total population of the $i$th deme, and $R$ be a standard distance scale (chosen by us), then we define the interaction matrix
$$
J_{ij} = \frac{ J N_j \exp\left(-\frac{1}{2 R^2} \Vert \bv{r}_i-\bv{r}_j\Vert^2\right)}{\sum_{j=1}^n N_j \exp\left(-\frac{1}{2 R^2} \Vert \bv{r}_i-\bv{r}_j\Vert^2\right)}.
$$
We will take $R=100$km  which is about twice the average deme radius when $n=10^3$, roughly equivalent to allowing interactions with nearest and next nearest neighbours, with more heavily populated demes exerting greater influence on their surroundings. Under this model of $\mat{J}$, spatial interactions in our state field depend on an effective diffusion coefficient
$$
D_{\tx{e}} = \frac{JR^2}{2}.
$$
To see this, let $A_i$ be the area of the $i$th deme and define the population density
$$
\rho_i = \frac{N_i}{A_i}.
$$
Introducing continuous fields $\tilde{\rho}(\bv{r}_i)$ and $\tilde{x}(\bv{r}_i,t)$ which interpolate smoothly between the discrete values of $\rho$ and $X$ defined at the centroid of each deme, we have
\begin{align*}
\sum_{j=1}^n & \rho_j x_j(t) \exp\left(-\frac{1}{2 R^2} \Vert \bv{r}_i-\bv{r}_j\Vert^2\right) \\   
&\approx \int_{\RR^2} \tilde{\rho}(\bv{r}_j) \tilde{x}(\bv{r}_j,t) \exp\left(-\frac{1}{2 R^2} \Vert \bv{r}_i-\bv{r}_j\Vert^2\right) d\bv{r}_j \\
&\approx 2\pi R^2 \tilde{\rho}(\bv{r}_i) \tilde{x}(\bv{r}_i,t) +\pi R^4 \nabla^2 (\tilde{\rho}(\bv{r}_i) \tilde{x}(\bv{r}_i,t)).
\end{align*}
Similarly
$$
\sum_{j=1}^n \rho_j \exp\left(-\frac{1}{2 R^2} \Vert \bv{r}_i-\bv{r}_j\Vert^2\right) \approx 2\pi R^2  \tilde{\rho}(\bv{r}_i) +\pi R^4 \nabla^2 \tilde{\rho}(\bv{r}_i) 
$$
If we now assume that $R^2 |\nabla^2 \tilde{\rho}|/\tilde{\rho} \ll 1$ so that the population density of each deme is approximately equal to the average of its neighbours, then the interaction term may be approximated as
\begin{equation}
\sum_{j=1}^n J_{ij}(X_j-X_i) \approx \frac{J R^2}{2} \frac{\nabla^2(\tilde{\rho} \tilde{x})}{\tilde{\rho}} = D_{\tx{e}} \frac{\nabla^2(\tilde{\rho} \tilde{x})}{\tilde{\rho}}.    
\label{eqn:dff_apx}
\end{equation}
From this we see that our interaction model induces diffusive mixing of states, modulated by variations in population density. The typical time scale over which demes that are separated by a distance $L$ interact, is given by the \textit{diffusion time}
$$
t_{\tx{diff}} = \frac{L^2}{D_{\tx{e}}}.
$$
Since  $R$ is approximately twice the deme radius, then $J^{-1}$ may be interpreted as the time (in years) for interactions to transmit across a single deme.

\section{Application to USA dialect data}
\label{sec:USA}

\begin{figure}
 \centering
        \includegraphics[width=1\columnwidth]{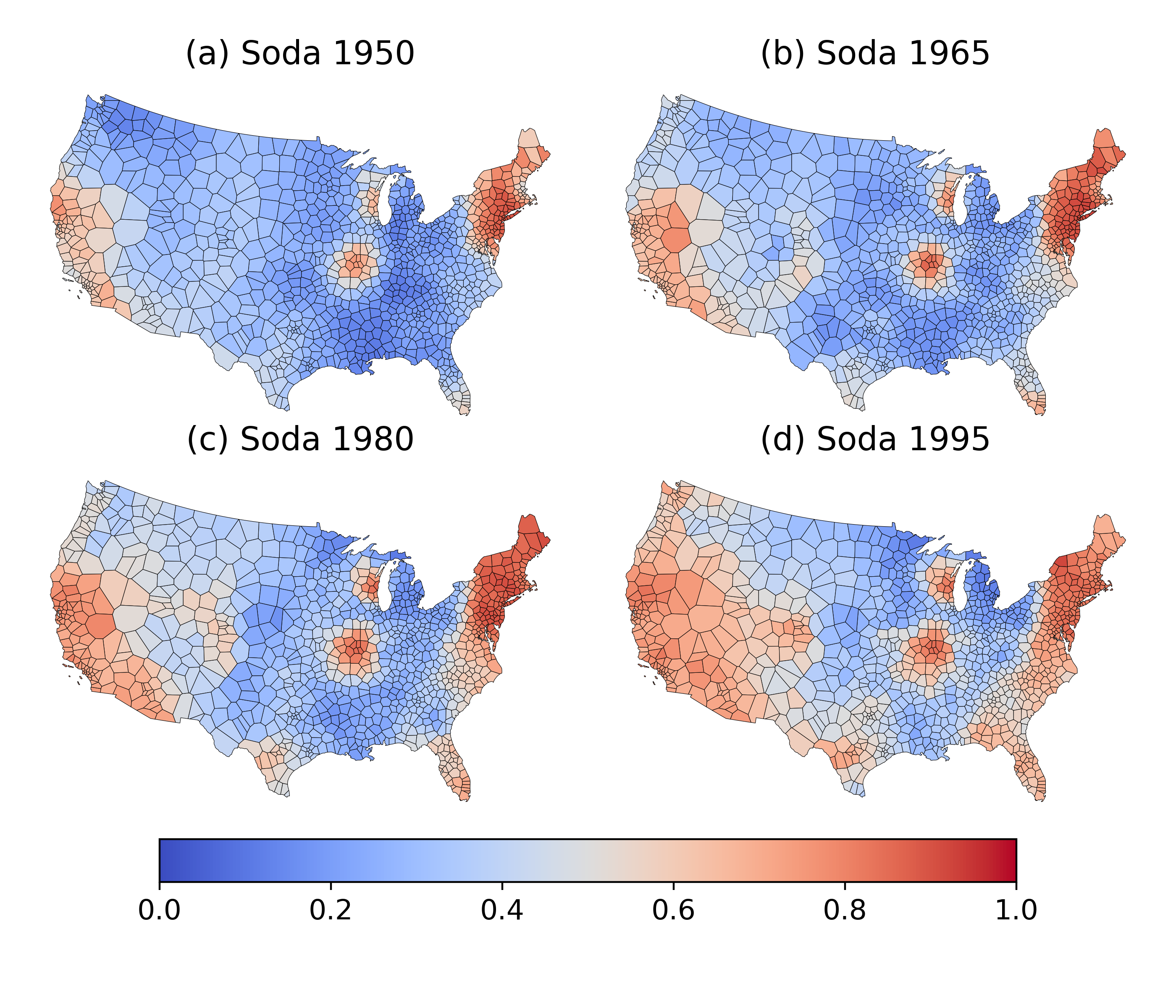}
 \caption{ Fractions of speakers (for different birth years) who use the `soda' variant, estimated using local logistic regression. }
\label{fig:soda}
\end{figure}

\begin{figure}
 \centering
        \includegraphics[width=1\columnwidth]{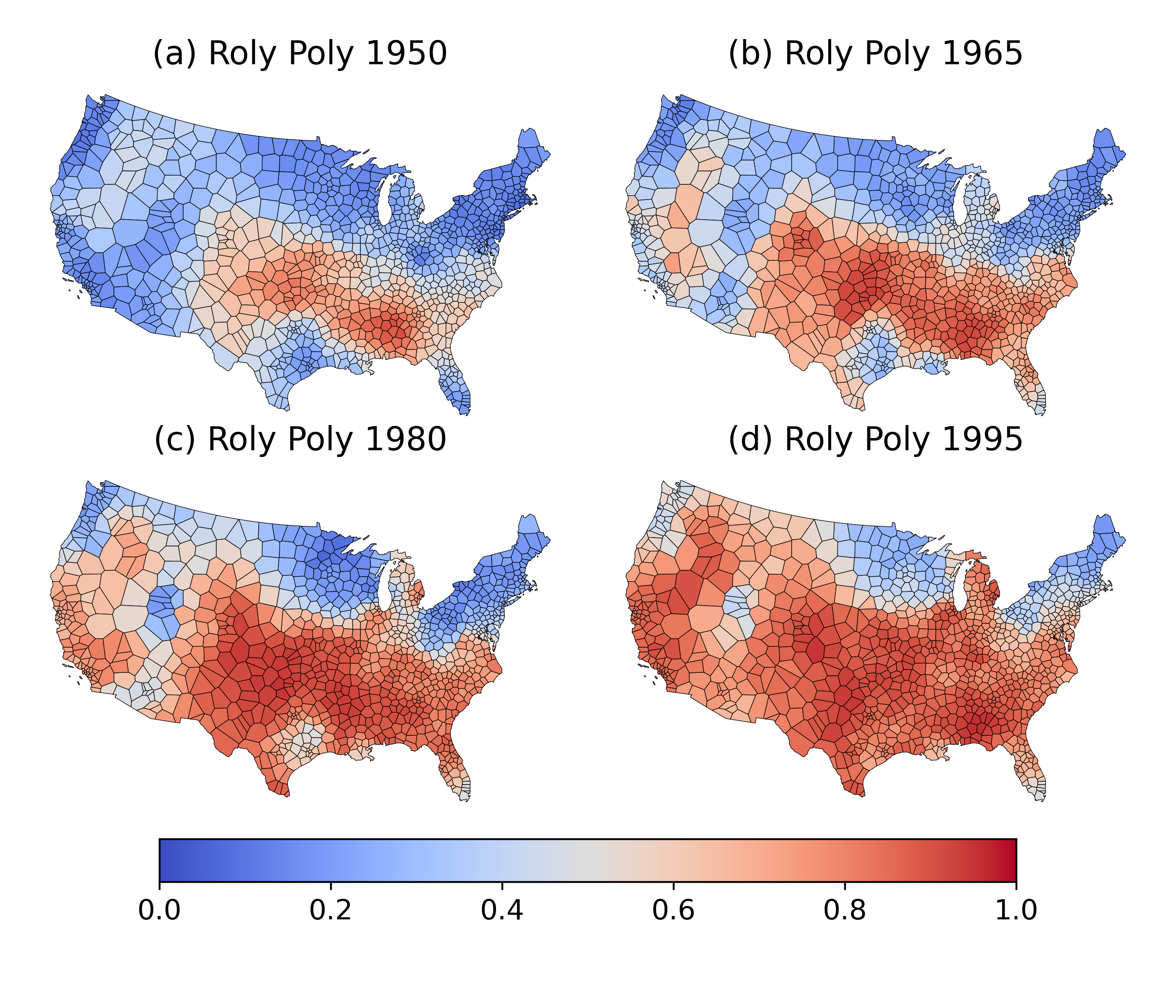}
 \caption{ Fractions of speakers (for different birth years) who use the `roly poly' variant, estimated using local logistic regression. }
\label{fig:roly}
\end{figure}

To explore our methodology we infer model parameters for two different linguistic features which have undergone substantial changes during the latter half of the 20th century in the USA. These data were collected as part of the `Cambridge Online Survey of World Englishes' ($\approx 10^5$ particiants) \cite{vau00}, initiated by Bert Vaux in 2007, and still active at time of writing.  The two linguistic variables we consider were surveyed via the the following questions.
\begin{enumerate}
\item What is your generic casual or informal term for a sweetened carbonated beverage?
    \item What do you call the little gray (or black or brown) creature (that looks like an insect but is actually a crustacean) that rolls up into a ball when you touch it?
\end{enumerate}

The most common answer to the first question was `soda' (41\%), with other commonly reported variants being `pop' (17\%), `coke' (10\%) and `soft drink' (10\%). The most common answer to the second question was `roly poly' (43\%) with other common variants being `pillbug' (15\%) and `potato bug' (11\%). 
Both questions elicited several further variants with popularities $\leq 5\%$. We refer to the variables (as opposed to the variants) as \textit{soda-pop} and \textit{woodlouse}. For each variable, we will consider only the most popular single variants --- soda and roly poly --- grouping together the others as `not soda' and `not roly poly'.  Survey respondents (`speakers') were also asked to provide  a zip code giving the location where they acquired their dialect features, and their date of birth (the apparent time variable). We use `time',  `apparent time' and `birth date' synonymously.

\subsection{Local logistic regression}

Before fitting our model we estimate the state fields at different apparent times via local logistic regression \cite{has09}. Considering a single variable, we let $v_i \in \{0,1\}$ be the variant choice of speaker $i$, and define the location-time-variant dataset
$$
D = \{((\bv{z}_i, t_i),v_i)\}_{i=1}^N
$$
where $\bv{z}_i$ and $t_i$ are the location and birth date of the $i$th speaker.  Introducing a reference time $t$ we define space and time displacements for the $i$th speaker with respect to $(\bv{r}_k,t)$ 
\begin{align*}
    \delta \bv{z}_{ki} &= \bv{z}_i - \bv{r}_k \\
    \delta t_i &= t_i - t.
\end{align*}
We then define the \textit{local} dataset
\begin{align*}
D_k(t)& = \\
&\left\{(\bv{z}_i,t_i,v_i) \in D \tx{ s.t. } \Vert \delta \bv{z}_{ki} \Vert < \delta z_{\tx{max}} \wedge |\delta t_i|< \delta t_{\tx{max}} \right\}     
\end{align*}
where $\delta z_{\tx{max}}=200$km and $\delta t_{\tx{max}}=10$ years. We model the state field in the vicinity of $(\bv{z}_k,t)$ using the logistic model
$$
\bv{x}(\bv{z}_k + \delta \bv{z} ,t+\delta t) = \tx{expit}\left(\beta_0 + \bv{\beta}_1^T \delta \bv{z} + \delta \bv{z}^T \mat{\beta}_2 \delta \bv{z} + \alpha \delta t \right)
$$
where $\tx{expit}(u)=(1+e^{-u})^{-1}$. In this way we allow local variations in the log-odds of variant $A$ which are quadratic in space and linear in time. Having fitted this model to the local dataset, we estimate the state field at the deme centroid as
$$
\hat{\bv{x}}_k(t) = \tx{expit}(\hat{\beta}_0).
$$
The procedure is then repeated for all demes over a sequence of reference times of our choosing. For some demes in sparsely populated areas we find only a relatively small number of datapoints ($<200$). In such cases we increase $\delta z_{\tx{max}}$ to the point where $|D_k(t)|>200$. Figures \ref{fig:soda} and \ref{fig:roly} show the time evolution of the soda and roly poly variants over the second half of the 20th century, estimated by the procedure.

\begin{figure}
 \centering
        \includegraphics[width=1\columnwidth]{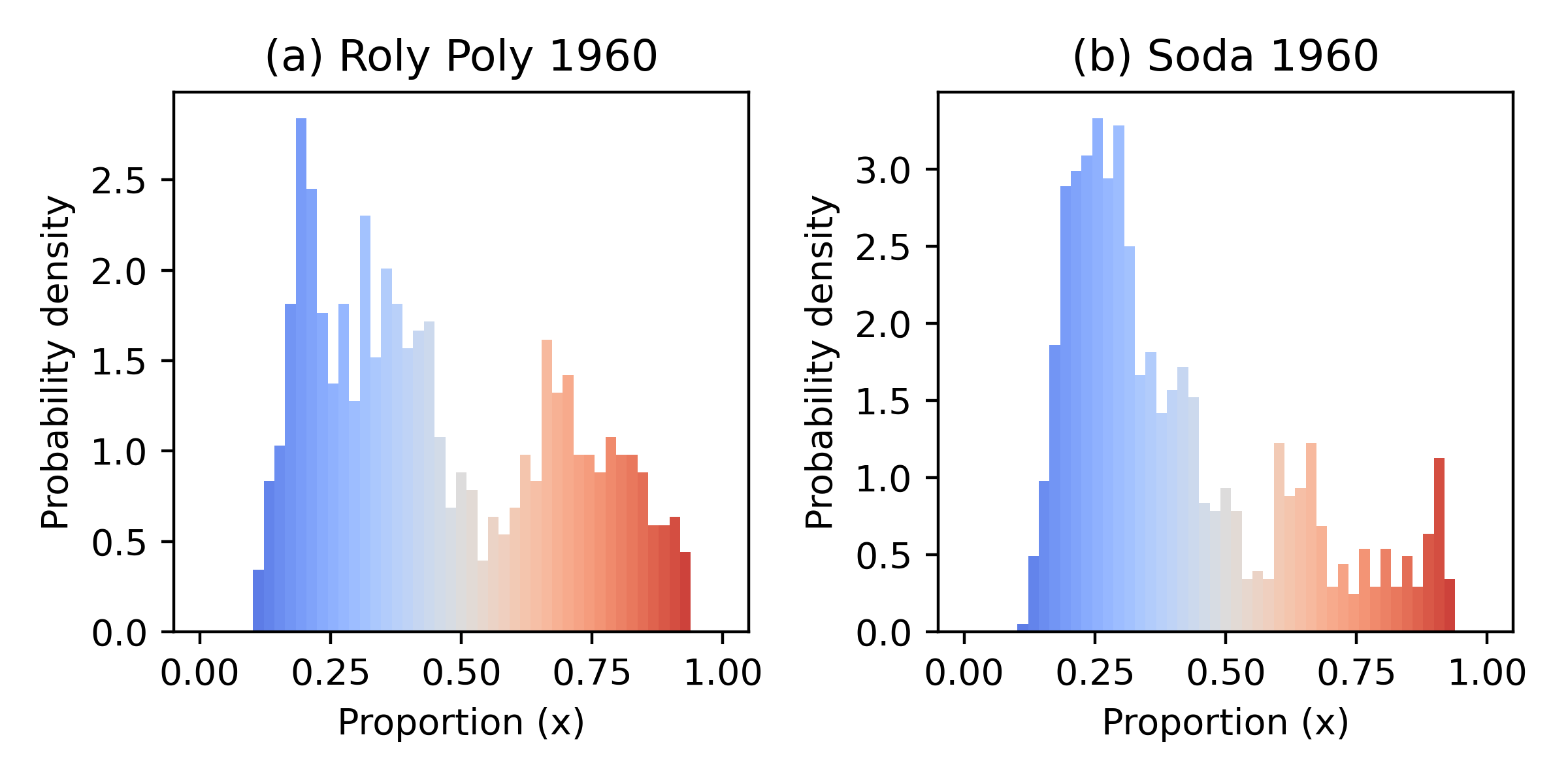}
 \caption{ Probability distributions of the fraction of speakers using the Roly Poly and Soda variants across all demes in 1960.  }
\label{fig:bimodal}
\end{figure}

We make the following observations about Figures \ref{fig:soda} and \ref{fig:roly}. First, the state field distributions are approximately bimodal --- locally, we typically see either a substantial majority of speakers using the $A$ variant (soda or roly poly)  or a substantial majority using the $a$ variant. This bimodality may be observed more clearly in Figure \ref{fig:bimodal} which shows the probability distribution of variant proportions across all demes in 1960. Although not visible in Figure \ref{fig:soda}, usage of the second most popular variant --- pop --- is also highly localised (in Michigan, Indiana and Ohio). Our second observation is that the boundaries of the regions where one variant dominates are quite narrow in comparison to the regions themselves (see also Figure \ref{fig:soda_zoom}). This suggests the presence of conformity driven dynamics. Our third observation is that the $A$-variant regions have expanded substantially over the survey period, indicating the presence of bias.

\subsection{Inference test}

\begin{figure}
 \centering
        \includegraphics[width=1\columnwidth]{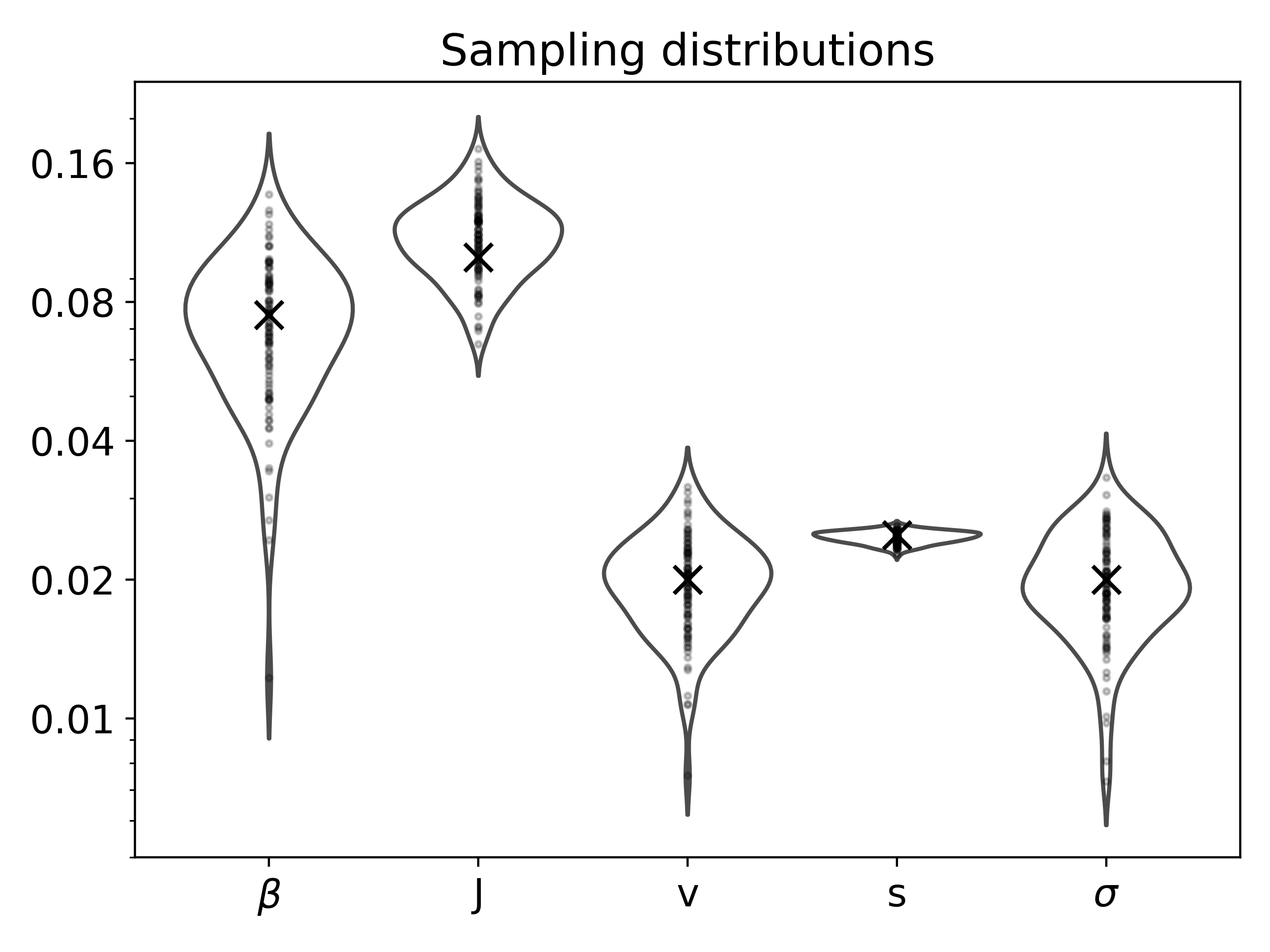}
 \caption{ Approximate sampling distributions (100 samples) of parameters inferred from simulations of a synthetic model. For each simulation, $7 \times 10^3$ synthetic responses were generated (matching the number in the real data). True parameter values, sample means and standard deviations are $\beta = 0.075$ ($\hat{\beta} = 0.073 \pm 0.024$), $J=0.1$ ($\hat{J} =0.11 \pm 0.021$), $\nu=0.02$ ($\hat{\nu}=0.020 \pm 0.005$), $s=0.025 $ ($\hat{s}=0.025\pm 0.001$), $\sigma=0.02$ ($\hat{\sigma} =  0.020 \pm 0.005$). True parameter values are shown as crosses inside the violins. }
\label{fig:sampling}
\end{figure}

Before fitting to survey data we test our inference procedures by simulating the model using parameters of our choosing, sampling a set of synthetic observations of the same size as our real dataset, then estimating the model parameters from these observations. We use as our initial condition the local regression estimate of the  roly poly state field in 1950 (Figure \ref{fig:roly}a), and simulate for a period of 50 years using a constant bias field. By repeating the simulation, data generation and inference process multiple times we are able to estimate the sampling distribution of our parameter estimates. The marginal sampling distributions of each parameter are shown as violin plots in Figure \ref{fig:sampling} (violin width is proportional to probability density). True parameter values, sample means, and standard deviations are provided in the caption. From  Figure \ref{fig:sampling} we see that the bias estimate is the most accurate, and that estimators other than the interaction strength are close to unbiased. The interaction strength estimator is biased weakly downwards, but lies well within the sample standard deviation. We therefore expect deviations in the estimate from the true value to be predominantly due to noise.

\subsection{Time independent bias field}

\label{sec:t_ind}

We now consider the model in the limit $\tau \ra \infty$ with $\kappa^2 \sim \tau^{3/2}$, so that the bias field is constant in time (there is an infinite prior penalty on time fluctuations) but there is no prior regularisation of its magnitude. In this case, the flexibility of the field is entirely controlled by its spatial length scale $\eta$. We will infer parameters by initialising the state field for each variable to our local regression estimate for 1950, then repeatedly solving forward to 2000, adjusting parameters to maximise the quasi posterior. Birth years in the data are concentrated in the interval $[1950,2000]$ (93\% lie within it) with on average 1420 respondents born in each year. Observations before 1950 are sparsely distributed in apparent time, and there is very little data available after 2000. 

\begin{figure}
 \centering
        \includegraphics[width=1\columnwidth]{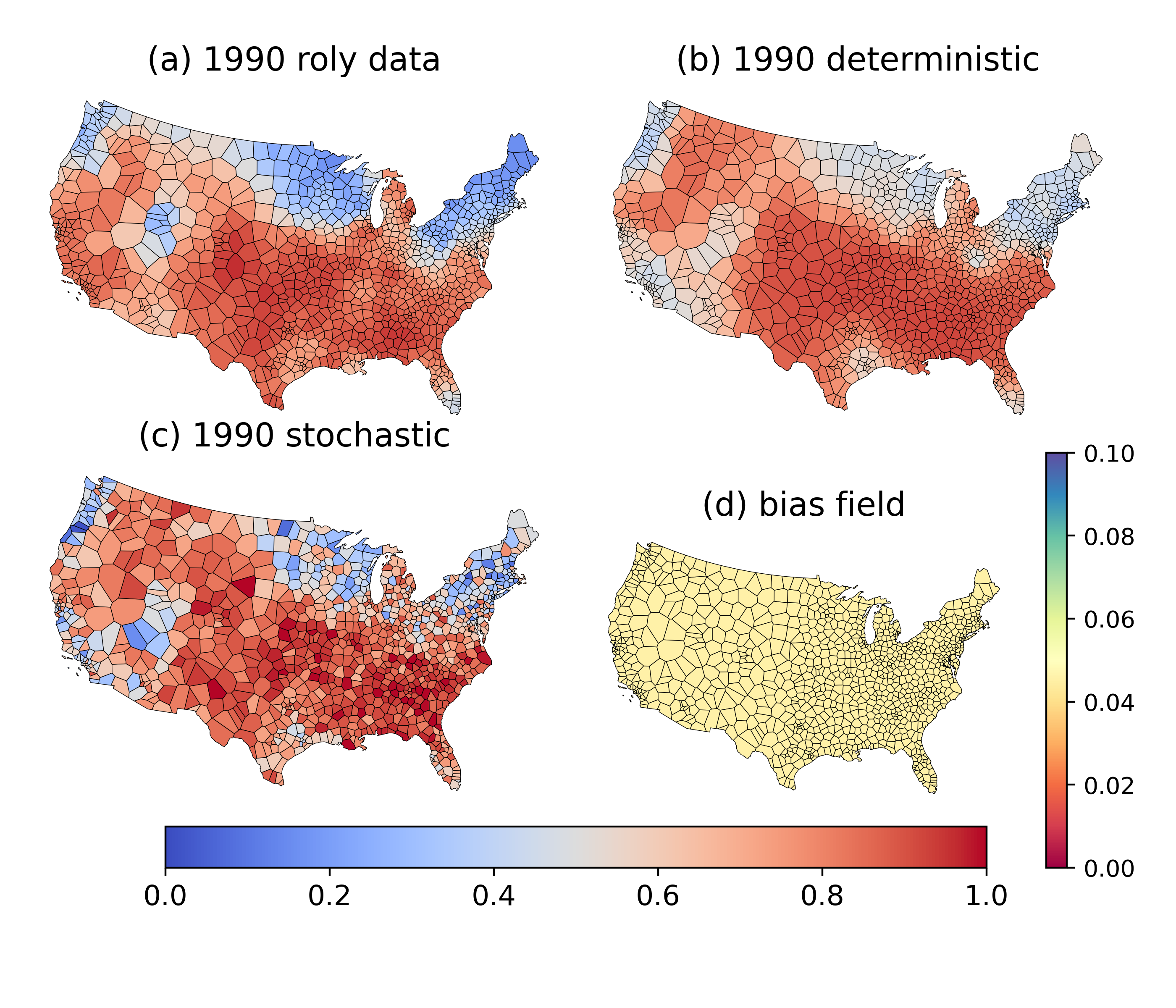}
 \caption{ (a) Local regression estimate of the roly poly fraction in 1990. (b) Deterministic (zero noise) solution to inferred model with $\eta=10^5$km. (c) Simulation of inferred stochastic model. (d) Bias field. Inferred parameter values are $\hat{\beta} = 0.053$, $\hat{J}=0.012$, $\hat{\nu}=0.007$, $\hat{\sigma}=0.027$.  }
\label{fig:big_eta}
\end{figure}

Consider first the case where the spatial length scale is very large ($\eta = 10^5$km) so that the bias field is effectively constant in space as well as time. To illustrate the behaviour of the fitted model in this case, in Figure \ref{fig:big_eta} we plot, for 1990, the the local regression estimate of the state field, the deterministic (zero noise) solution to the inferred model, and a sample from the full stochastic inferred model. The bias field is also shown in Figure \ref{fig:big_eta}; in this case it is constant. Values of inferred parameters are given in the caption. From Figure \ref{fig:big_eta} we see that the spatial expansion of the roly poly variant observed in the data is captured broadly by the model, but with some notable discrepancies. Two key differences are that in the data, Los Angeles and San Francisco experienced a dramatic rise in roly roly usage during the study period (see Figure \ref{fig:roly}), and that roly poly usage has remained stable in the Northeast and northern Midwest in the data, but in the model it has declined  in this region. We will refer to these effects as \textit{California expansion} and \textit{Northeastern stability}. We observe that Salt Lake City in the western USA is a hold-out area against the spread of roly poly, more so in the data than in the model.

An explanation for these discrepancies between model and data is as follows. When the bias field is constant in space and time, there is only one \textit{universal} bias parameter. The inferred mutation parameter is very small, and has little effect of spatial patterns, leaving spatial interactions, conformity and noise to explain all changes. To capture the overall expansion of the roly poly variant, a positive bias is needed. In the presence of bias, high conformity, which drives variant usage toward the currently most popular variant, is needed for Northeastern stability. However, the same conformity effect will oppose California expansion. The two observations cannot simultaneously be generated by the model.  We hypothesise that the inferred conformity parameter is unrealistically low due to this tension, and that the inferred noise is unrealistically high --- the model is forced to describe systematic discrepancies as fluctuations.

\begin{figure}
 \centering
        \includegraphics[width=1\columnwidth]{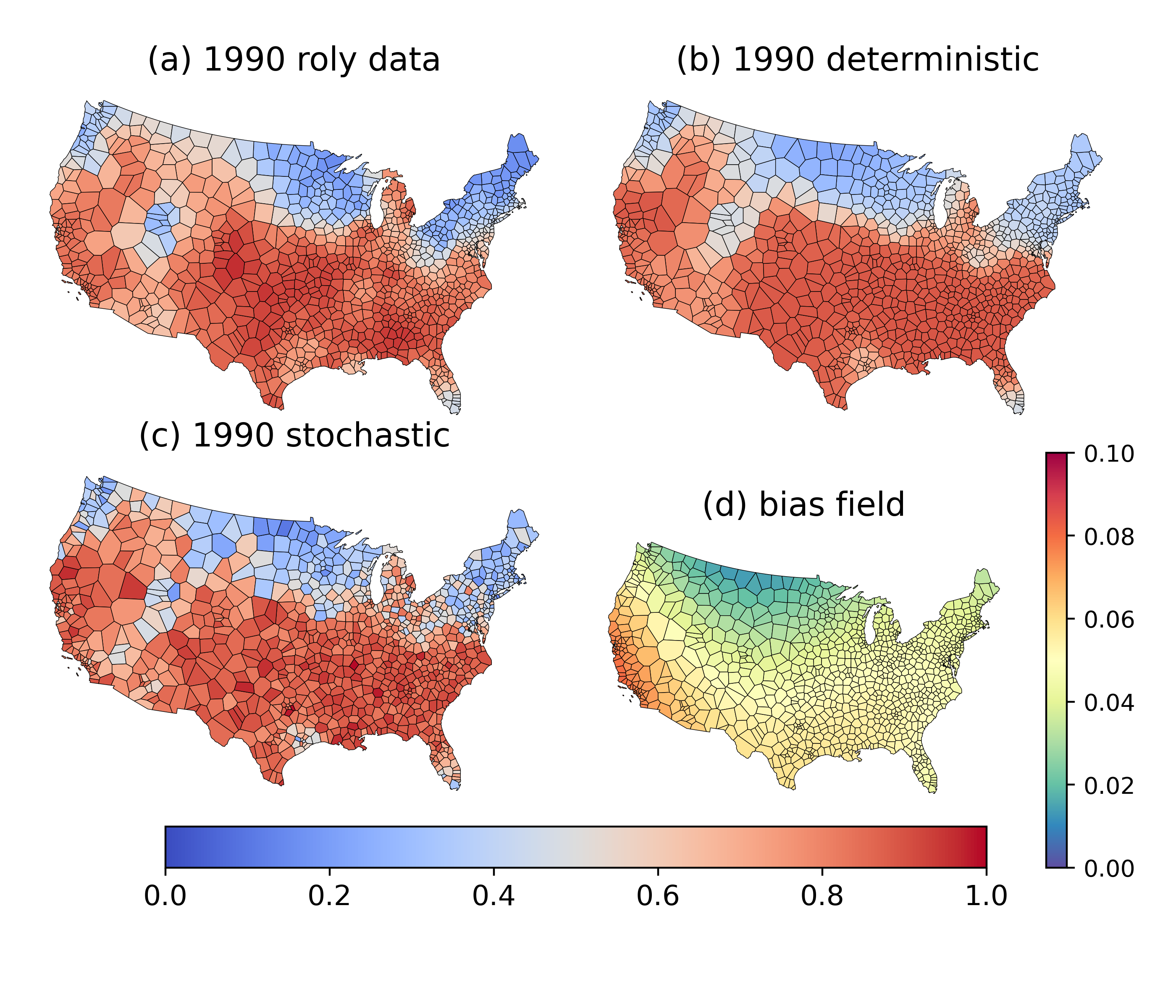}
 \caption{ (a) Local regression estimate of the roly poly fraction in 1990. (b) Deterministic (zero noise) solution to inferred model with $\eta=1.5 \times 10^3$km. (c) Simulation of inferred stochastic model. (d) Bias field. Inferred parameter values are $\hat{\beta} = 0.125$, $\hat{J}=0.020$, $\hat{\nu}=0.018$, $\hat{\sigma}=0.022$.  }
\label{fig:medium_eta}
\end{figure}

We now reduce the bias field length scale to $\eta=1500$km. In this case, an $m=8$ component bias field retains 99\% of field variance (see section \ref{sec:lifting}). Figure \ref{fig:medium_eta} shows the same comparisons that were made in Figure \ref{fig:big_eta}. Here the bias field (Figure \ref{fig:medium_eta}c) has approximately the same background value, but with a `hot-spot' in California, and a `cold-spot' along the Canadian border (but not noticeably in Northeastern USA). With this increase in model complexity, the previously noted discrepancies largely disappear. The inferred values of other parameters also change (see caption in Figure \ref{fig:medium_eta}). In particular conformity has more than doubled, spatial interactions have increased and noise reduced. Due to the California bias hot-spot, the conformity effect can be overcome in this region, allowing the parameter to play a role in explaining Northeastern stability. The increase in conformity has also notably \textit{sharpened} the interfaces of the regions in which roly poly dominates, making them more consistent with observations.

It is not surprising that allowing extra flexibility in the bias field has improved the ability of the model to explain observations. It is therefore important to consider whether the model is telling us anything useful about the \textit{physics} of language change. As stated in section \ref{sec:bias}, we do not provide any explicit interpretation of the bias field, or explanation of its origin. What, then, can we learn form its form? The first thing we learn is \textit{where} and to \textit{what extent} non-neutrality needs to be injected into the model. If a relatively simple field accurately captures observed evolution, and the inferred parameters describing neutral, translation-invariant aspects are large enough to significantly affect the model behaviour, this suggests they are part of the physics of language change. Moreover, the effects induced by these parameters --- surface tension driven coarsening, order-disorder transitions --- have close analogues in physical systems. This opens up the possibility that the machinery developed to describe physical systems using statistical field theory can be adapted to explain aspects of the evolution of language. The form of the inferred bias field also shows linguists and dialectologists where `surprising' changes have taken place --- beyond those encoded in neural processes. In the current example it is clear that something unusual has happened in California to induce a switch to the roly poly variant. The final value of the bias field is practical; it allows us to `fill the gaps' in the neutral model, allowing it to match observations in a minimalist way. Even if we don't have an explicit explanation for the particular form taken by the field, the hope is that it is a reasonable proxy for the real mechanisms at play. In this case it will provide a means to predict future changes. The time scale over which we can predict the future will depend on how quickly the underlying mechanisms change. We will return to this point in section \ref{sec:future}. 

\begin{figure}
 \centering
        \includegraphics[width=1\columnwidth]{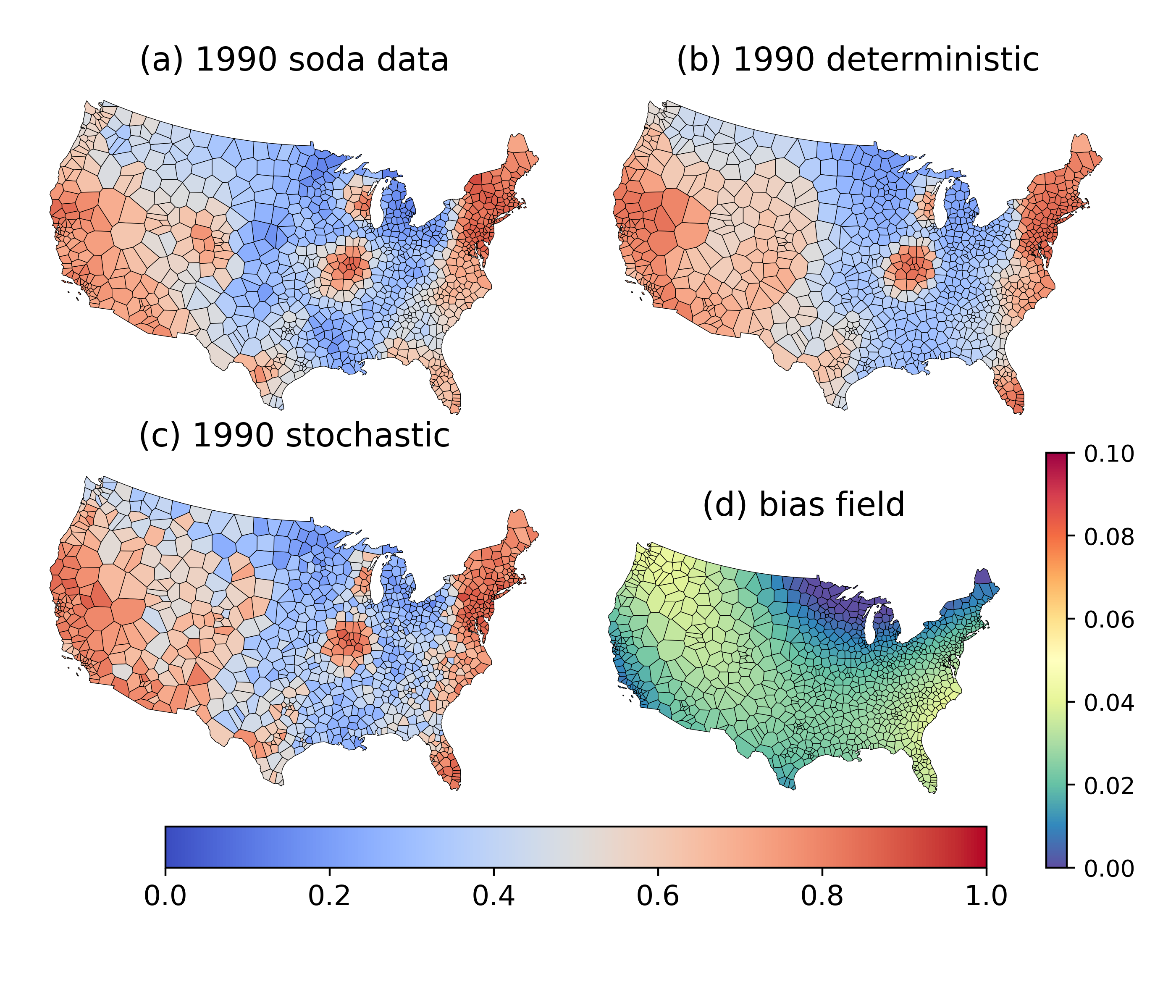}
 \caption{ (a) Local regression estimate of the soda fraction in 1990. (b) Deterministic (zero noise) solution to inferred model with $\eta=1.5 \times 10^3$km. (c) Simulation of inferred stochastic model. (d) Bias field. Inferred parameter values are $\hat{\beta} = 0.106$, $\hat{J}=0.014$, $\hat{\nu}=0.017$, $\hat{\sigma}=0.012$.   }
\label{fig:soda_medium_eta}
\end{figure}

Now consider our second linguistic variable --- the word for a sweet fizzy drink. We have repeated the fitting process for the $\eta=1500$km bias field, with the results for 1990 shown in Figure \ref{fig:soda_medium_eta}.  Here again the model is able to closely reproduce the observed evolution of the soda variant. The inferred conformity parameter takes a similar value to the roly poly case (the values differ by less than the standard deviation of the sampling distribution obtained in our inference test). In contrast, the inferred noise variable for soda is approximately half the value for roly poly. This is consistent with the observation that less commonly used variables experience faster variant evolution \cite{pag07} (`soda' is $\approx 10^3$ times more common on Google Ngram than `roly poly' \cite{aid10}). Our final observation regarding the evolution of soda is that the cities of St. Louis and Milwaukee represented two isolated `islands' of soda usage. The bias in favour of soda is relatively low in these regions (particularly in Milwaukee, where it is approximately zero). In this situation, by analogy with the `shrinking droplet' effect in non-conservative phase ordering \cite{bra02}, we would expect these islands to shrink and disappear due to surface tension at their boundaries. It has been hypothesised \cite{bur17} that they don't because population gradients induce an effective force on the interface, pushing it toward lower density regions. We explore this hypothesis in more detail in section \ref{sec:tension}.

\subsection{Primordial soup test}

\begin{figure}
 \centering
        \includegraphics[width=1\columnwidth]{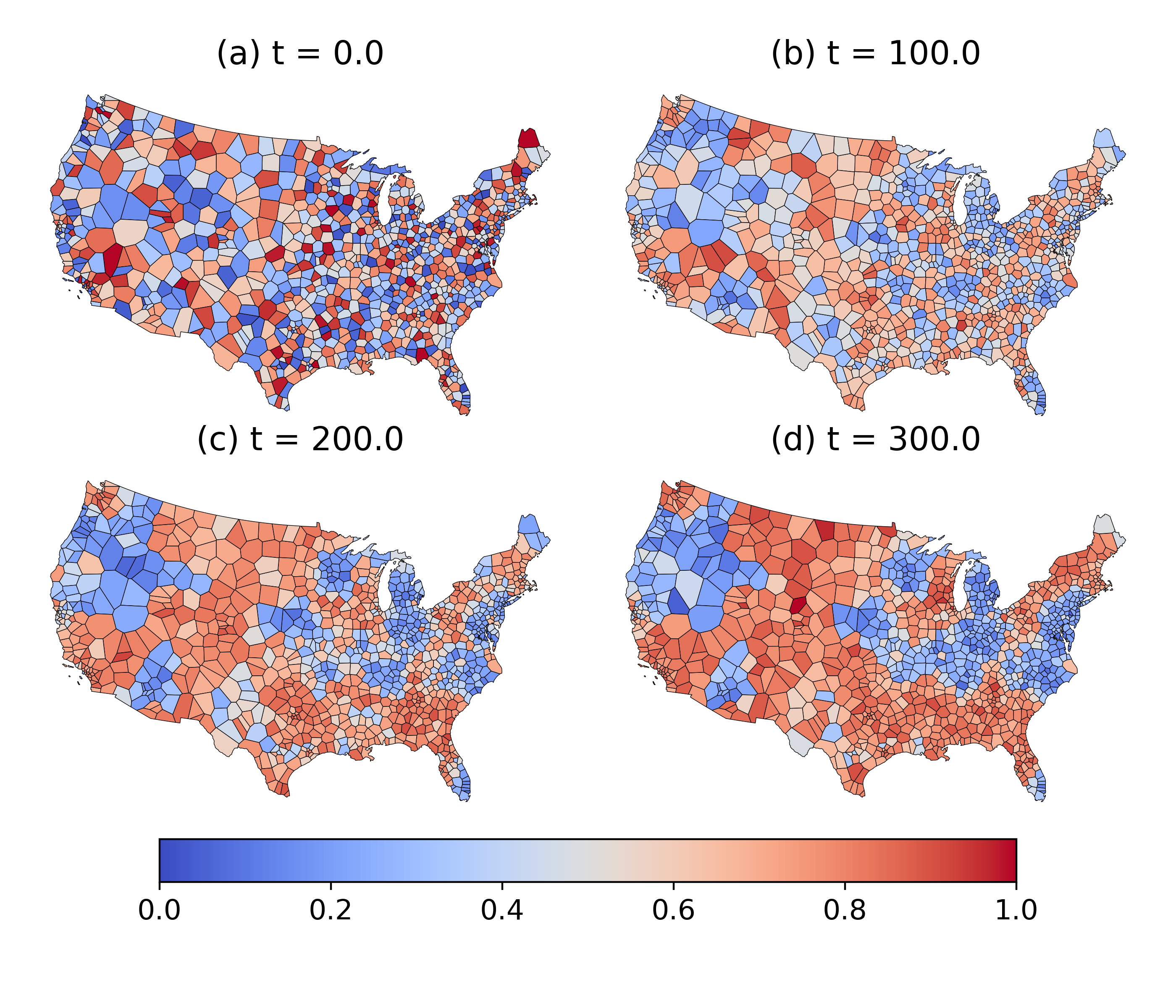}
 \caption{Evolution of zero bias model from `primordial soup' conditions, using parameters inferred from the woodlouse variable (Figure \ref{fig:medium_eta}).   }
\label{fig:primordial}
\end{figure}

Having fitted our model to two examples of variant evolution, we now address the question of whether the parameter estimates we obtained are consistent with observed patterns of variation more generally. The examples we have considered are typical of the geographical pattern of variation seen in many different linguistic features. They consist of regions where single variants dominate, with the interfaces between regions having well defined boundaries (isoglosses). Historical surveys of dialect variation show that the sizes of domains in which single variants dominate have gradually increased over time. For example, in the UK the pattern of dialect has significantly coarsened since the 19th century  \cite{mag12, bri09}. In the examples above the model was initialised with a state field that already exhibited substantial domains dominated by a single variant. Do such patterns emerge spontaneously from our dynamics?  To address this question we initialise our model using a `primordial soup' of variant usage, where the variant frequency in each deme is generated uniformly at random from the interval $[0,1]$. We then evolve this state using the parameters inferred from the woodlouse variable, but with the bias field set to zero. In this case, evolution is driven entirely by noise, conformity and mutation. Figure \ref{fig:primordial} (a) shows the primordial state, with panels (b)-(d) showing later states, separated by one hundred year intervals. We see that over time, large single-variant domains emerge. 

We hypothesise that as with phase ordering in physical systems, the coarsening observed in Figure \ref{fig:primordial}  is driven by a surface-tension effect which tends to smooth out domain boundaries \cite{bra02}. In the linguistic case the the motion of these boundaries will also be influenced by variations in population density. An effect of this kind has previously been explored for a deterministic spatial model of language change \cite{bur17}, in which domain walls experience an effective force which drives them towards regions of lower population density. We derive the form of this force for the current model in section \ref{sec:tension}. Inspection of Figure \ref{fig:primordial} (d) reveals that domains boundaries often surround densely populated metropolitan areas, consistent with this effect. We address this point analytically in section \ref{sec:tension}. 

\begin{figure}
 \centering
        \includegraphics[width=1\columnwidth]{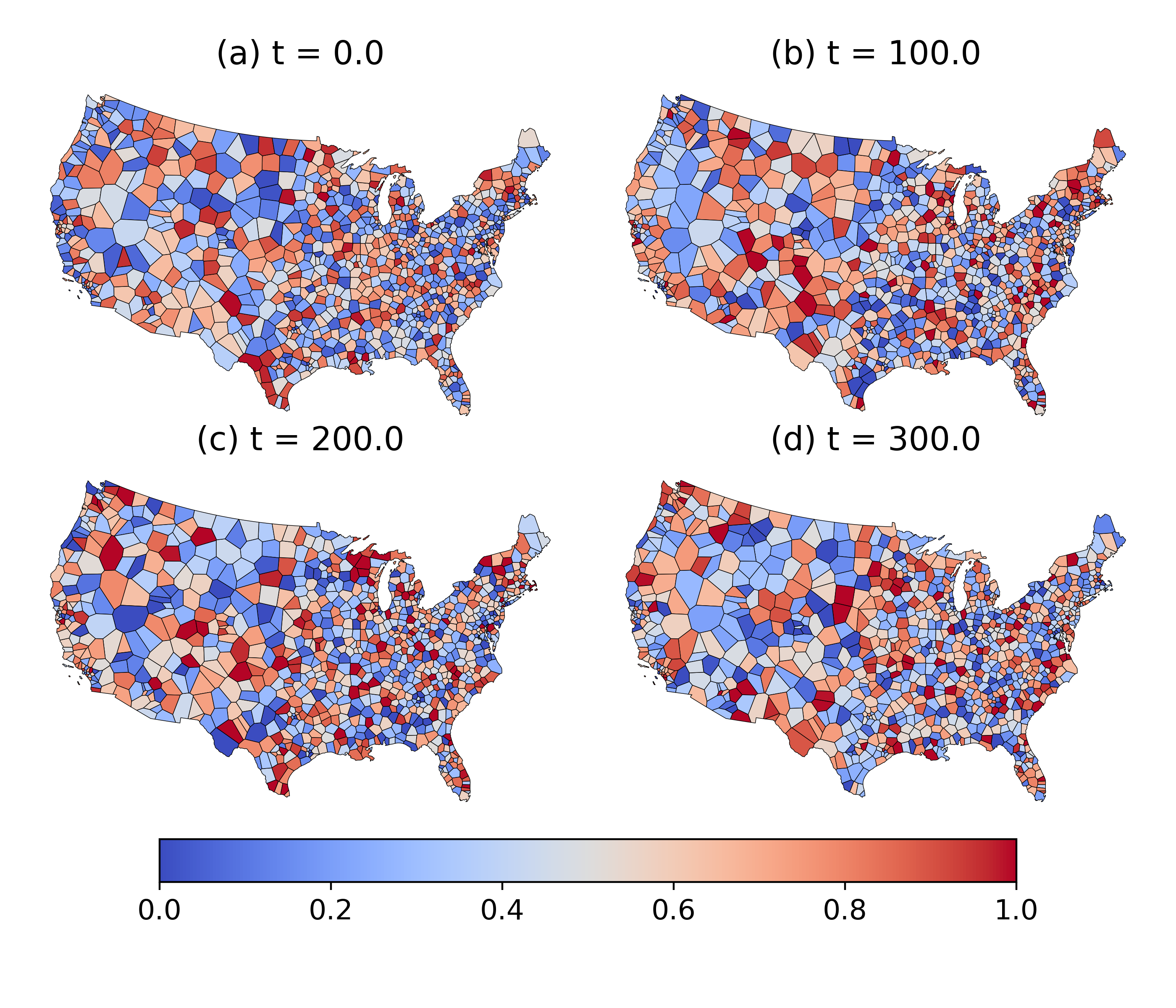}
 \caption{Evolution of zero bias model from `primordial soup' conditions using $\sigma=0.075$, with other parameters inferred from the woodlouse variable (Figure \ref{fig:medium_eta}).  }
\label{fig:primordial_disorder}
\end{figure}

Single-variant domains are not the only kind of spatial pattern observed in language surveys. We also see examples of highly mixed states with little long range order \cite{vau00}. The coarsening effect shown in Figure \ref{fig:primordial} is present only when the noise parameter is sufficiently small.  Increasing $\sigma$ beyond a critical value, depending on the other model parameters, we see an order-disorder transition. In the disordered regime, primordial initial conditions do not coarsen. An example is given in Figure \ref{fig:primordial_disorder}. 

\subsection{Time varying bias}

\begin{figure}
 \centering
        \includegraphics[width=1\columnwidth]{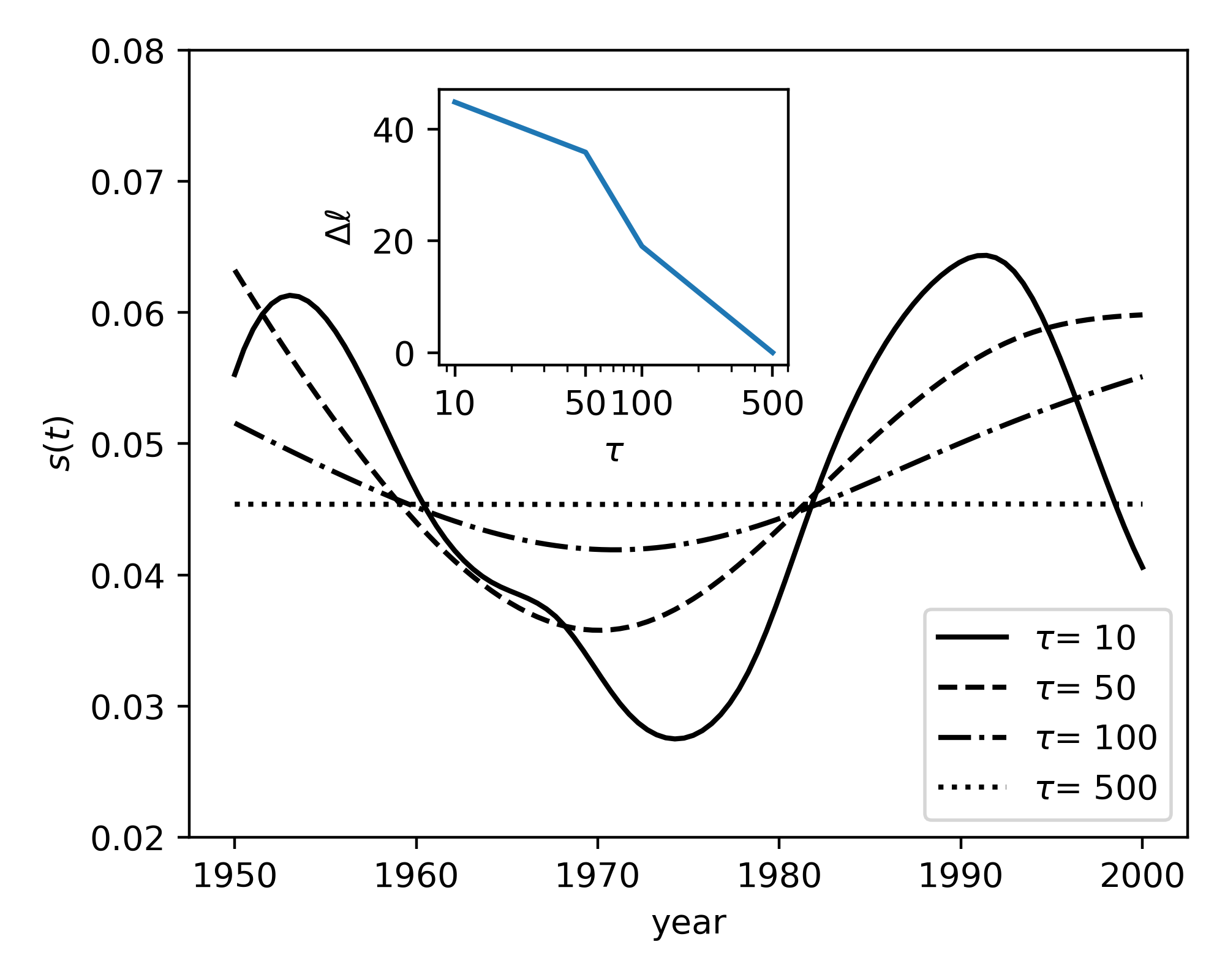}
 \caption{ Inferred bias fields for the woodlouse variable over the interval $[1950,2000]$. Main plots show inferred fields for different fluctuation time scales, and inset plot shows difference in log likelihood function relative to the long-timescale case $\tau=500$ years.    }
\label{fig:s_tau}
\end{figure}

We now consider the behaviour of the model when the bias field is allowed to fluctuate over finite time scales. We consider the case where spatial fluctuations are absent ($\eta \ra \infty$). Combined space-time fluctuations are considered in section \ref{sec:future}. To explore the influence of the time scale $\tau$, in Figure \ref{fig:s_tau} we have obtained MAP estimates of time varying bias fields for the woodlouse variable over the period $[1950,2000]$. These estimates were obtained by representing the embedded field as a cubic spline \cite{has09} interpolating between field values at a sequence of fixed time points or `knots', with knot spacing chosen to be substantially smaller than $\tau$. The field values at the knot points were then determined simultaneously with the other state field parameters by numerical optimisation. We set $\kappa=1$, weakly regularising the magnitude of fluctuations in the embedded field, and producing average bias field values close to those obtained in time independent case (section \ref{sec:t_ind}) where $\kappa \ra \infty$.

As we might expect, reducing $\tau$ produces fields which fluctuate more rapidly in time. The fact that the model `makes use' of the extra flexibility provided by shortening the time scale implies that the fit to data is improved by reducing $\tau$. This is confirmed by the inset plot in Figure \ref{fig:s_tau}, which shows the difference, $\Delta \ell$, in the log-likelihood relative the longest time scale fit.  The question of what is the most appropriate choice for $\tau$ may be addressed using two different approaches. One is to compare the likelihood gain produced by adding flexibility, with the increase in the effective number of parameters added to the model by doing so. Metrics used for this purpose, designed to minimise overfitting by balancing the trade-off between model fit and model complexity, include Mallow's $C_p$ statistic \cite{mal73} and the Akaike Information Criterion \cite{aka74}. A second approach --- the one we adopt --- is to use $\tau$ to maximise the predictive ability of the model. We do so by viewing $\tau$ as controlling the time scale over which knowledge of the past is useful as a means to predict the future. 

\section{Bias field half life and future prediction}

\label{sec:future}

\begin{figure}
 \centering
        \includegraphics[width=1\columnwidth]{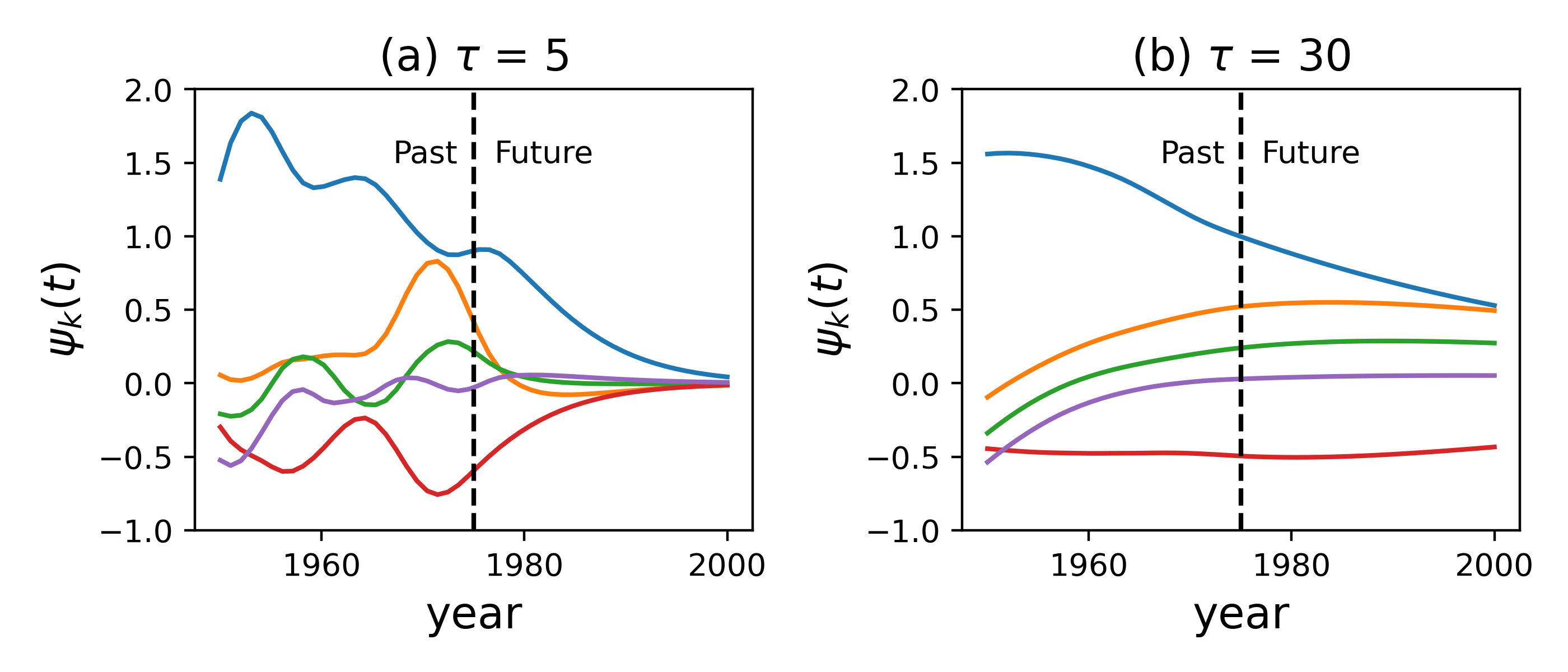}
 \caption{ Components of the embedded bias field ($\eta=1500$km, $R_\ast^2=95\%)$ fitted to woodlouse observations from $[1950,1975]$ then evolved forward in time by solving the Euler Lagrange equations associated with the Onsager-Machlup action. }
\label{fig:past_future}
\end{figure}

The inference procedures developed in section \ref{sec:inf} allow us to obtain MAP estimates for the bias field over any time period, $[t_0,t_1]$, for which observational data is available. These estimates balance the competing objectives of maximising the likelihood and minimising the Onsager-Machlup (OM) action (\ref{eqn:SOM}), which penalises paths that are implausible according to the bias field model given by equations (\ref{eqn:S}--\ref{eqn:V}). Outside of the data interval ($t>t_1$) the most plausible embedded bias fields minimise the OM action subject to boundary conditions at  $t=t_1$. Such fields  are solutions to the corresponding Euler Lagrange equation
$$
\frac{d^4\psi}{dt^4} - \frac{2}{\tau^2} \frac{d^2 \psi}{dt^2} + \frac{1}{\tau^4} \psi = 0.
$$
Here we have omitted component subscripts on $\psi$ for brevity. The general solution to this ODE is
$$
\psi(t) = (A+Bt)\exp\left(-\frac{t}{\tau}\right) + (C + Dt) \exp\left(\frac{t}{\tau}\right).
$$
These two terms may be understood as describing the behaviour of the field as $t \ra \pm \infty$. We are interested in the future behaviour of the field so we discard the second term. Solving subject to boundary conditions on the field and its first derivative at $t=t_1$ we have, for $t \geq t_1$, 
\begin{align}
\nonumber
\psi(t) = & \left(\left(1+ \frac{t-t_1}{\tau}\right)\psi(t_1) +  \left(t-t_1\right)\dot{\psi}(t_1)\right) \\
&\times \exp \left(-\frac{t-t_1}{\tau} \right).    
\label{eqn:ELsol}
\end{align}
To predict the future behaviour of the state field we first fit the embedded bias field within the data interval, then use (\ref{eqn:ELsol}) to extend the field into the future. Figure \ref{fig:past_future} shows the result for two different values of $\tau$, taking our data interval as $[1950,1975]$, then predicting forward to the year 2000. Whatever the value of $\tau$, the embedded field components will eventually decay to zero, with $\tau$ controlling time scale over which this takes place. The decay time also depends on the derivative of the field at $t_1$ --- if a component is moving away from the origin at the end of the data interval, it will continue to do so for some time, before beginning to return.  It is useful to define the `half-life' of the field via
$$
\psi(t_1+t_{\frac{1}{2}}) = \frac{\psi(t_1)}{2}.
$$
If $\dot{\psi}(t_1)=0$ then $t_{\frac{1}{2}} \approx 1.68 \tau$. Notice that due to the momentum of the field, this is longer than the half life of purely exponential decay ($t_{\frac{1}{2}}^{\tx{exp}}=\tau \log 2 \approx 0.69 \tau$) by a factor of $\approx 2.4$. It is clear from Figure \ref{fig:past_future} that the choice of $\tau$ can have a substantial impact on future predictions. In Figure \ref{fig:past_future} (a), where $\tau=5$, all components have decayed nearly to zero by the end of the century, 25 years after the end of the data interval. In Figure \ref{fig:past_future} (b), where $\tau=30$,  the field components have decayed very little by this point. 

\begin{figure}
 \centering
        \includegraphics[width=1\columnwidth]{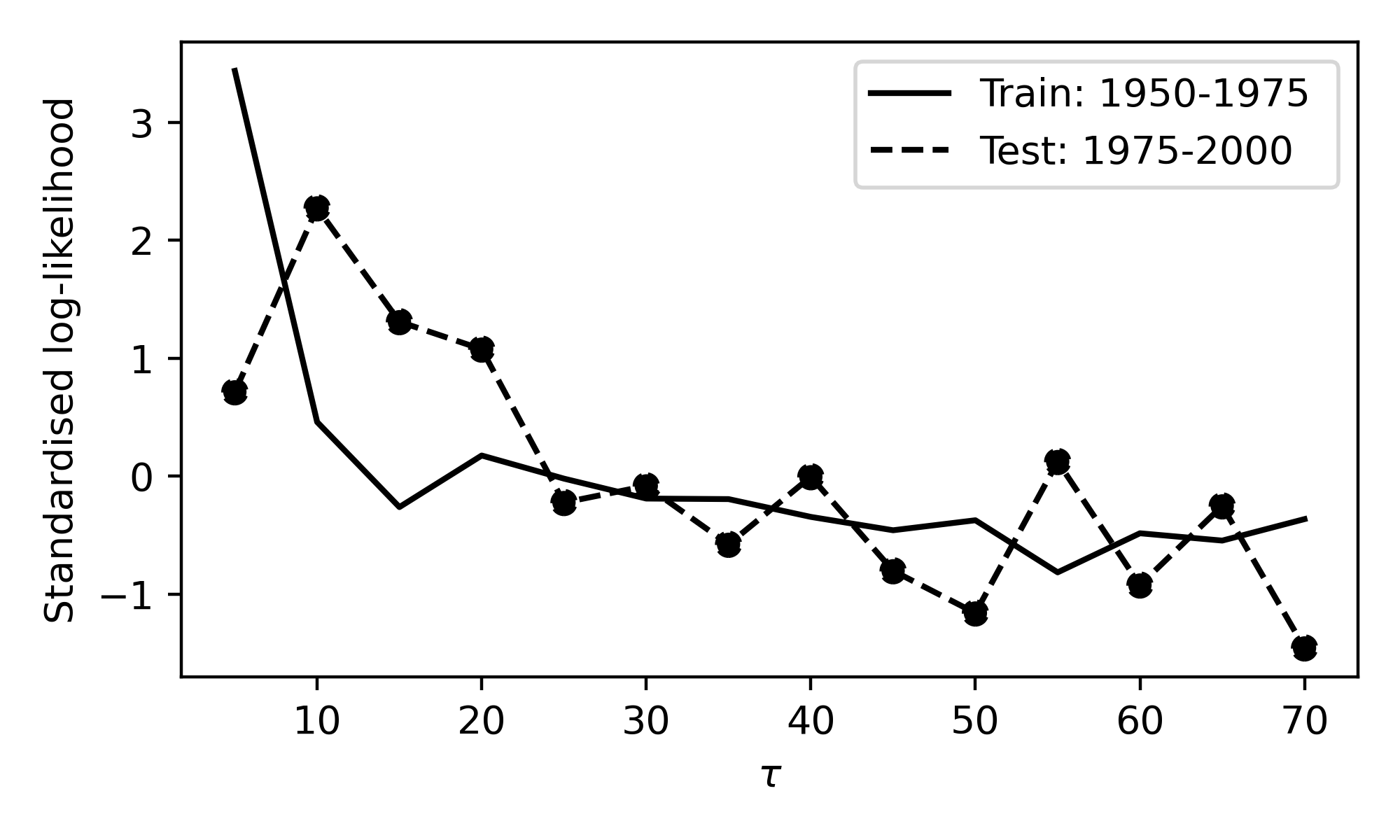}
 \caption{Standardised likelihoods evaluated on the data interval used for fitting (training) the model [1950,1975] and on a subsequent test interval [1975,200]. To standardise likelihoods were rescaled to have zero mean and unit variance when averaged over $\tau$. }
\label{fig:cv}
\end{figure}

\begin{figure}
 \centering
        \includegraphics[width=1\columnwidth]{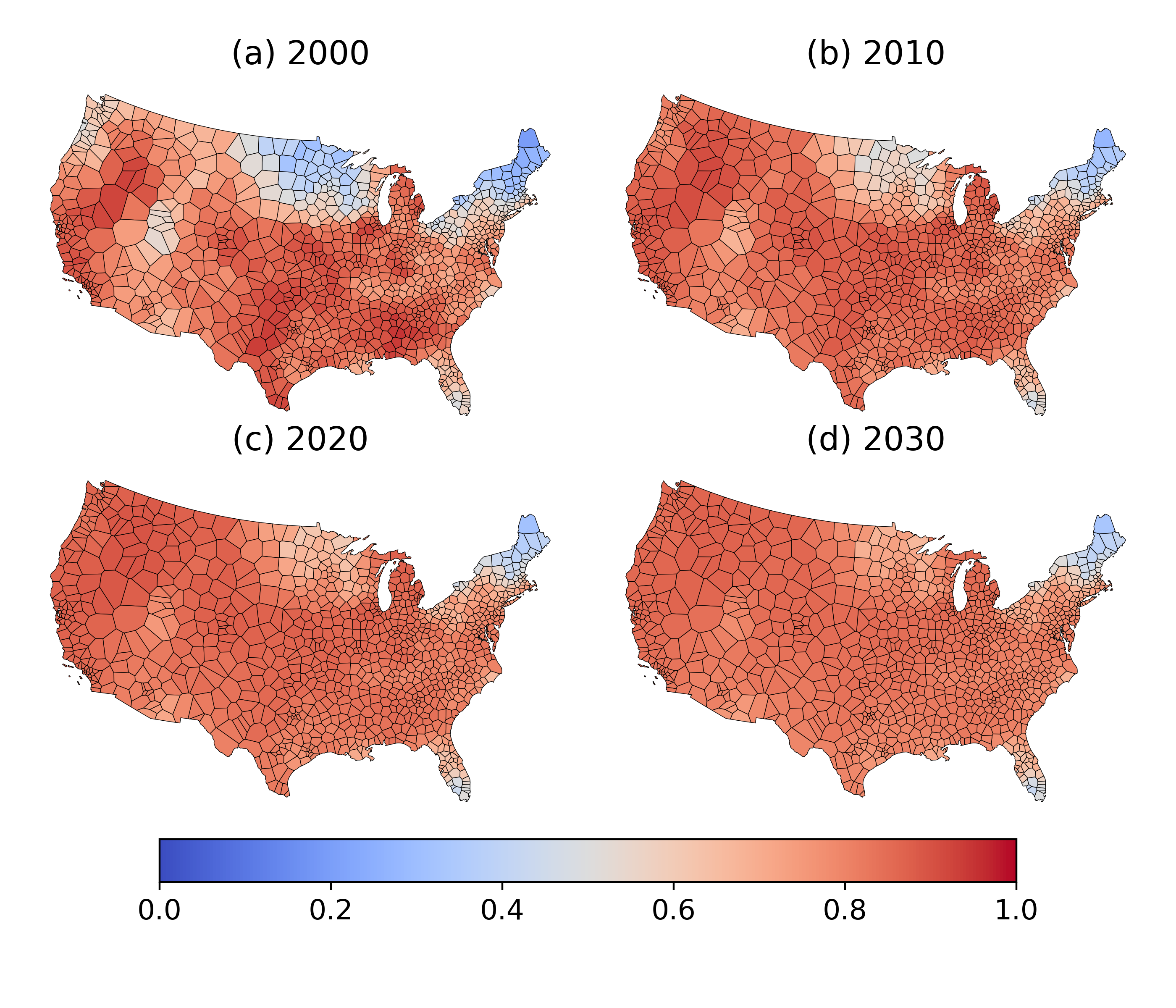}
 \caption{ Forward prediction of the roly poly fraction using $\tau^\ast=10$ with parameters inferred from [1980,2000]. }
\label{fig:roly_future}
\end{figure}

Although we do not expect to be able to predict the long term future of any given linguistic variable, we do expect recent behaviour of the bias field to be a guide to  its near term future. The parameter $\tau$ controls the time scale over which current field values, and trends in those values, persist. We can estimate its value by cross validation. We first split our observation interval into two halves. Observations in the first half [1950,1975] define the \textit{training data} $D_{\tx{train}}$, and observations in the second half [1975,200] the \textit{test data} $D_{\tx{test}}$. For a series of increasing values of $\tau$, we fit the model to the training data, then use the  parameter estimates obtained to forward predict bias and state fields into the test interval.  We then compute the cross entropy (log likelihood) of both the training and the test data using the fitted (and forward predicted) model. This performance measure is defined
\begin{align*}
&\ell(\tau) = \\
& \sum_{(t,y_k(t)) \in D} y_k(t) \log x_k(t) + (n_k(t)-y_k(t)) \log (1-x_k(t))   
\end{align*}
where $D \in \{D_{\tx{train}}, D_{\tx{test}}\}$. Figure \ref{fig:cv} displays standardised versions of this measure. Whereas the training measure declines systematically with increasing $\tau$ the test performance initially increases with $\tau$ before later declining.  Optimal performance on test data is obtained using $\tau^\ast=10$, corresponding to a half life of $\approx 17$ years. The intuition here is that if we set $\tau < \tau^\ast$ we will not fully exploit the information available from the current bias field, but if we set $\tau>\tau^\ast$ we will be projecting current trends too far into the future. 

\begin{figure}
 \centering
        \includegraphics[width=1\columnwidth]{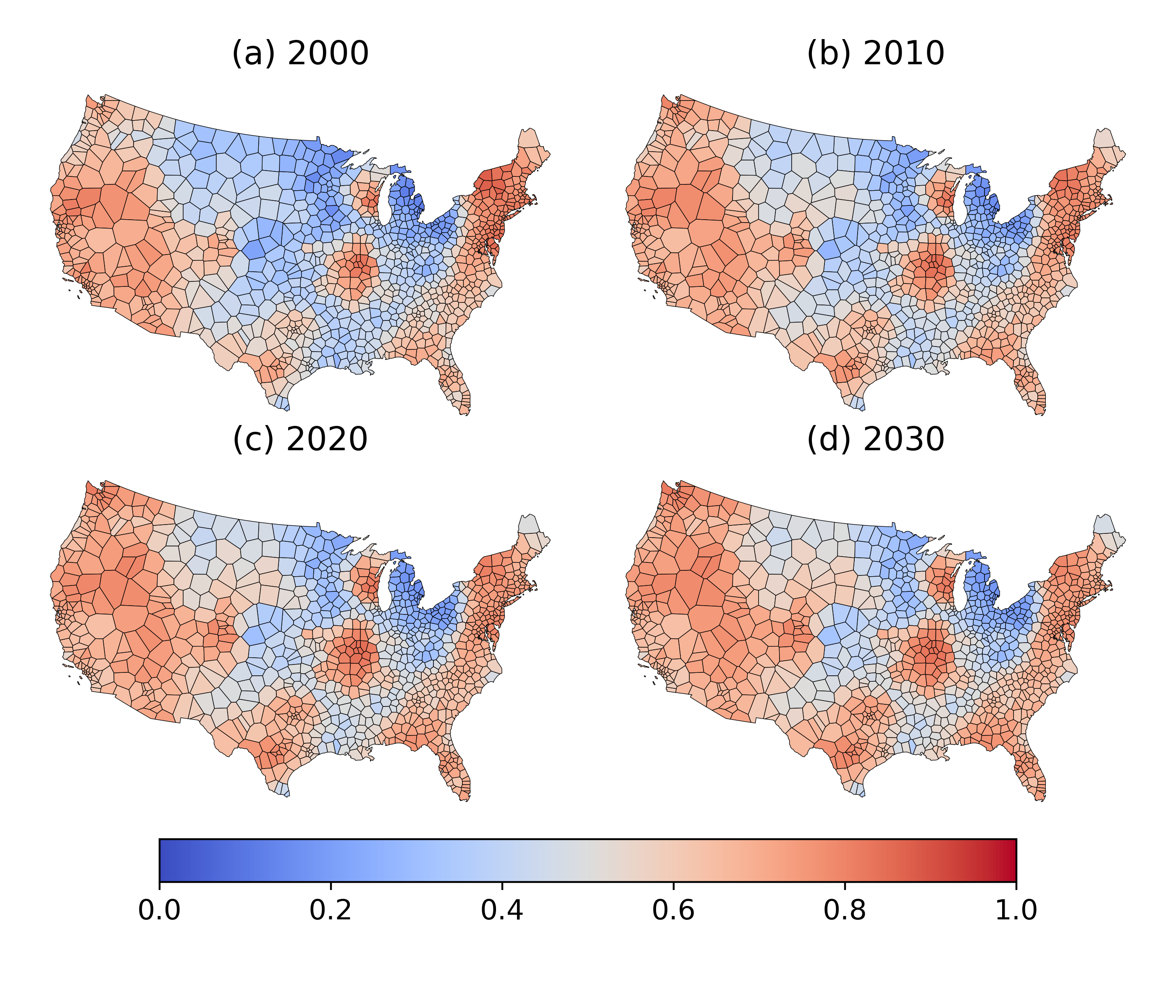}
 \caption{ Forward prediction of the soda fraction using $\tau^\ast=10$ with parameters inferred from [1980,2000].  }
\label{fig:soda_future}
\end{figure}

Having estimated $\tau$ we can now make future predictions. We first fit the model to the recent past [1980,2000], in order to obtain parameter estimates and the current state of the bias field. We then use our local logistic regression estimate of the state field at $t=2000$ as the initial condition for our model, and solve forward in time using MAP estimates for the future of the bias field. The results for the woodlouse and soda-pop variables are shown in Figures \ref{fig:roly_future} and \ref{fig:soda_future}. Since these are estimates for the states of speakers born in the years 2000-2030, they may be interpreted as predictions for the state of the language community as a whole, approximately 20 years ahead. For example, we predict that roly poly will become dominant throughout the USA except in Maine, Vermont and Miami by around 2040. Of course it it possible that a new and more popular variant will emerge in the meantime.

\section{Conformity, surface tension and population gradients}

\label{sec:tension}

In the examples we have considered, the inferred parameter values resulted in a model capable of generating domains in which one variant is dominant, provided the noise parameter was not too large. We now explore the dynamics of domain evolution analytically by considering a continuum approximation to our dynamics in the small noise limit.

In the limit of large numbers of demes the state field becomes a  continuous function of position and time $x(\bv{r},t)$. Utilising our diffusion approximation (\ref{eqn:dff_apx}) to the interaction term, then in the absence of noise and bias the state field obeys the partial differential equation
\begin{equation}
\frac{\pa x}{\pa t} = \beta (2x-1)x(1-x) + \nu (1-2x) + D_{\tx{e}} \frac{\nabla^2 (\rho x)}{\rho}. 
\label{eqn:pde}    
\end{equation}
We will use this equation to explore the dynamics of interfaces between domains, denoting position $\bv{r}=(r_1,r_2)$.

\subsection{Stable interface}

When the population density is constant in space, and when $\beta > 4 \nu$, then equation (\ref{eqn:pde}) admits a stable equilibrium (time-independent) solution in the form of a infinite straight interface between two domains with bulk frequencies given by 
$$
x^{\pm} = \frac{1}{2}\left( 1 \pm \sqrt{1-\frac{4\nu}{\beta}}\right).
$$
These are stable fixed points of (\ref{eqn:pde}). There is a third frequency, $x=\tfrac{1}{2}$, for which the non-linear term in (\ref{eqn:pde}) is zero, but this not a stable fixed point. Let us assume that our interface lies along the $r_2$ axis so that our solution is independent of $r_2$. Let $x(r_1)$ be the frequency at position $r_1$, then in equilibrium, we have
\begin{equation}
D_{\tx{e}} x'' + \beta (2x-1)x(1-x) + \nu (1-2x) = 0.    
\label{eqn:interface}
\end{equation}
It may be shown (see appendix \ref{app:inter}) that a solution to (\ref{eqn:interface}) is
\begin{equation}
x(r_1) = \frac{1}{2}\left(1 + \sqrt{1 - \frac{4 \nu}{\beta}} \tanh \left( \frac{r_1}{2}  \sqrt{\frac{\beta-4\nu}{D_{\tx{e}}}} \right) \right).    
\label{eqn:stable}
\end{equation}
This represents a domain wall or `isogloss' at $r_1=0$ separating domains with bulk frequencies $x^{-}$ and $x^{+}$. We can define the width of the isogloss as the size of the region within which its linear approximation around $r_1=0$ lies in the interval $[x^-,x^+]$. This width is
$$
w(D_{\tx{e}},\beta,\nu) = \sqrt{\frac{D_{\tx{e}}}{\beta-4 \nu}} = \sqrt{\frac{D_{\tx{e}}}{\beta}} + O(\nu).
$$
For example, using the parameters inferred for the woodlouse variable in Figure \ref{fig:medium_eta} we obtain
\begin{align*}
    w &\approx 43 \tx{km} \\
    x^+& \approx 0.83 \\
    x^- & \approx 0.17.
\end{align*}
We obtain almost identical results using the parameters inferred for pop-soda (Figure \ref{fig:soda_medium_eta}); $w=43$km and $x^+=0.81$. Key observations here are that isoglosses are widened by increasing diffusion (stronger or longer-range interactions) and made narrower by stronger social conformity, which also pushes bulk frequencies closer to zero or one.

\subsection{Interface motion}

\begin{figure}
 \centering
        \includegraphics[width=1\columnwidth]{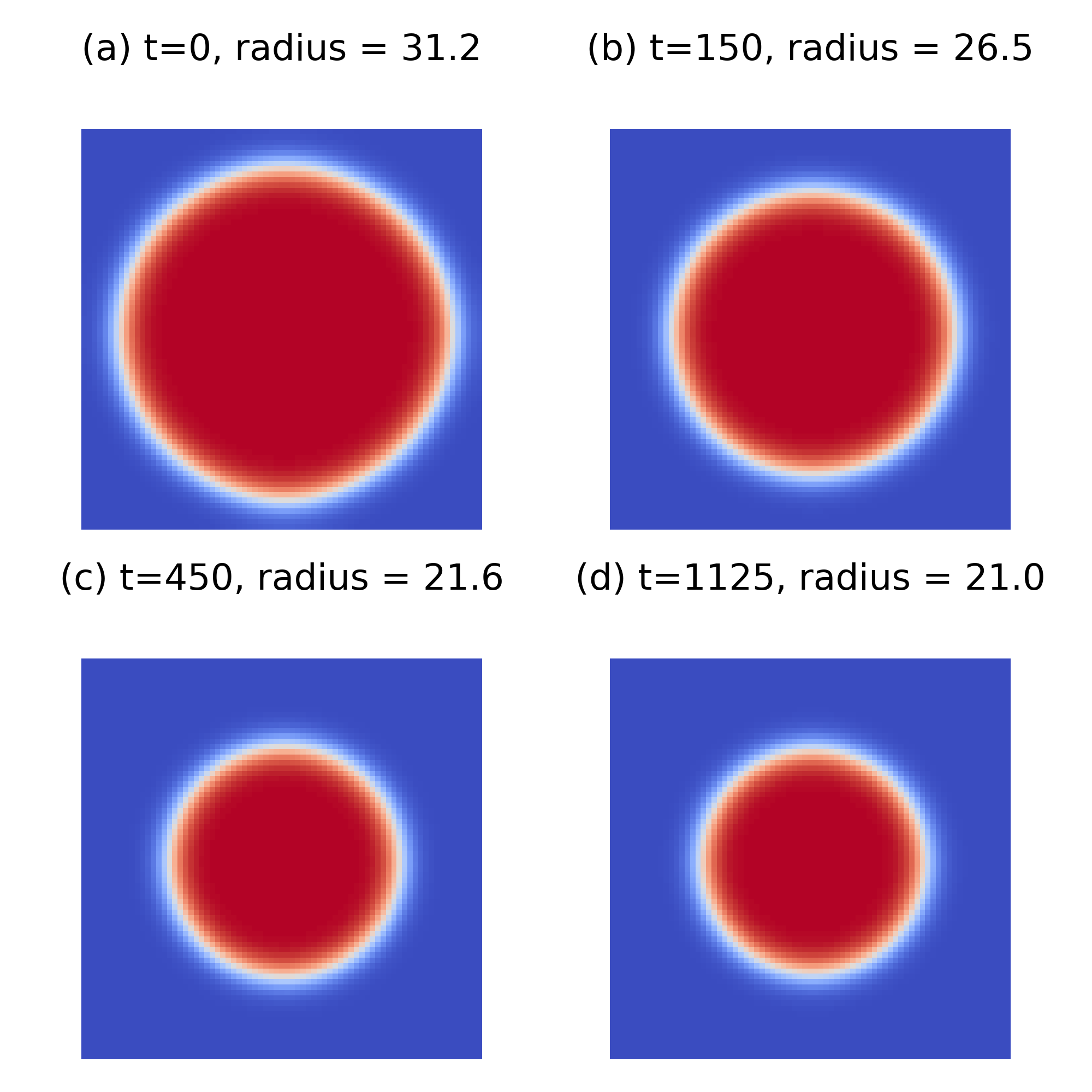}
 \caption{ Shrinking A phase droplet. Parameters $D_{\tx{e}}=1, \beta=0.5, \nu=0$.  Density field $\rho(r) = (1-\alpha) + \alpha \exp(-r^2/(2 r_0^2)$ with $\alpha=0.7$, $R=8$.    In this case $r_{\tx{eq}}=20.8$. }
\label{fig:droplet}
\end{figure}

Whereas straight interfaces with profile given by (\ref{eqn:stable}) are stable equilibria of the dynamics, curved interfaces are not. We can view an interface as the set of points for which $x(\bv{r})=\tfrac{1}{2}$. Given any point $P$ along an interface, if $\hat{\bv{g}}$ is a unit vector normal to the interface at $P$ then in the absence of bias the velocity of the interface in the direction of $\hat{\bv{g}}$ is given by
\begin{equation}
v \approx -D_{\tx{e}} \left( \nabla \cdot \hat{\bv{g}} + 2 \frac{\nabla \rho \cdot \hat{\bv{g}}}{\rho} \right). 
\label{eqn:v}
\end{equation}
This result is derived in appendix \ref{app:inter} (see also \cite{bur17} for a similar result in a memory-based language model).  The divergence of the unit normal to the interface is its curvature
$$
\kappa = \nabla \cdot \hat{\bv{g}} = \frac{1}{R}
$$
where $R$ is the radius of curvature of the interface at $P$. Intuitively, if the interface is highly curved then speakers growing up on the interface will be exposed to more of the variant lying on the side with positive curvature. They will therefore tend to adopt this variant, moving the interface in a direction which reduces its curvature. The population dependent contribution to (\ref{eqn:v}) causes interfaces to move in the direction of reducing population density. Intuitively, speakers growing up on an interface with more people on one side than the other will tend to adopt the variant used on the more densely populated side, moving the interface toward lower density regions.

To illustrate the combined effect of curvature and population density on interface motion, we consider a circularly symmetric city-like population density
$$
\rho(r) = (1-\alpha) + \alpha \exp(-r^2/(2r_0^2)
$$
where $r = \sqrt{r_1^2+r_2^2}$. We initialise the state field as a circular droplet, concentric with the city but with a much larger radius. Letting $r$ be the radius of the droplet then the velocity in the radial direction, from (\ref{eqn:v}), is
$$
v(r) = \frac{2 \alpha r}{r_0^2(\alpha + (1-\alpha) \exp(r^2/(2r_0^2))}-\frac{1}{r}.
$$
Well away from the city the population density is constant so interface evolution is purely curvature driven. In this case the curvature component of the velocity acts to shrink the droplet. As the the edge of the city is approached the population gradient effect repels the droplet, which reaches a stable equilibrium if the population gradient is sufficiently large. The stable droplet radius, $r_{\tx{eq}}$ is then the larger of the two roots of $v(r)=0$. Figure \ref{fig:droplet} shows and example of this evolution. We see a close match between the stable radius predicted analytically and numerically. 

The fact that population gradients at the boundaries of cities are capable of repelling isoglosses provides a possible explanation for the stability of the soda isogloss around St. Louis (see Figure \ref{fig:soda_zoom}) despite the fact that the bias in favour of the soda variant was very low.  A more thorough investigation of the implications of curvature and population gradient driven isogloss evolution may be found in \cite{bur17}.


\section{Conclusion}

We have introduced a statistical field model of language change in which a state field and a bias field evolve simultaneously in time. The state field dynamics --- a version of Wright-Fisher diffusion \cite{eth11} --- are derived from speaker-level processes which interpolate between biased voter and Ising-type dynamics within a spatially heterogenous population distribution. The dynamics of the bias field are controlled by parameters which restrict its freedom to fluctuate in space and time. We have developed efficient inference methods for learning model parameters from spatial data in `apparent time' \cite{san18}, and inferred our model parameters for two linguistic variables in the USA \cite{vau00}. We were able to reconstruct the history of our two variants, and our inferred parameter values indicate that surface tension driven coarsening has played in important role in this history. By estimating the `half-life' of the bias field, we were also able to make some future predictions. 

Our motivation for constructing this model is to take steps toward a `statistical field theory of language evolution'. Unlike statistical field theories of physical systems, a theory of language will not be able to fully explain the evolution of linguistic variables without resorting to exogenous factors. Language is inherently coupled to political, societal and technological change \cite{lab94,lab01,lab10,wei68,mil92,mil12}. While some changes in these categories may be predictable \cite{gol10}, many may not be \cite{pop57}. Our results suggest that despite this coupling, language evolution \textit{does} involve speaker level processes which lead to some macroscopic predictability. By inferring the parameters of these processes we are able to tease out the extent to which they are responsible for the changes we see, with  the bias field providing a minimalist representation of what extra driving processes are needed to explain observations. Further, because the processes in our model are so similar to those which drive physical systems, there is hope that the mathematical tools and physical insights that are available for those systems \cite{kar07,bin92} will be deployable in the context of language. In the current paper, for example, we have seen the importance of coarsening processes and stochasticity induced disorder.

There are many more steps to be taken in order to arrive at a satisfactory general theory.  First, we must consider more general state spaces. In the current paper we considered discrete (binary) variables. A natural next step is to allow more than two states, and to allow coupling between different variables. Many phonological variables (describing vowel sounds for example) are naturally continuous, requiring a continuous state variable for their description \cite{hil95}. A challenge when expanding state space in this way will be to avoid overfitting. As we increase  dimensionality we also increase the number of parameters needed to describe our dynamics, and therefore the quantity of data required to infer those parameters.  In this context Bayesian methods offer a promising approach, allowing us to use expert linguistic knowledge and diverse data sources to shrink the hypothesis space of models.   

To capture language evolution over longer periods we need to allow the possibility for new linguistic variants to spontaneously enter the model. New variants often emerge without obvious exogenous cause. A theoretical mechanism which allows this to take place is `linguistic momentum' according to which speakers respond to age gradients in language use, meaning that stochastic fluctuations can trigger large scale spontaneous shifts \cite{sta16,mit17}.

Beyond more sophisticated state spaces and dynamics, we must also consider the geographical or social space within which evolution takes place, and the connectivity of that space. In the current paper we considered a static population distribution and relatively short range interactions mediated purely by geographical proximity. A more realistic model should also account for social as well as  spatial groupings, and changes in population distribution over time \cite{wie11,fag10,fan18}. Again the threat of overfitting will be present. A promising approach, consistent with the difficulty of observing the details of interaction networks, is to describe connectivity using low dimensional latent (unobserved) variables.

\begin{acknowledgments}
    The author thanks Bert Vaux for insightful comments on this article and for permission to use woodlouse and pop-soda data from the Cambridge Online Survey of World Englishes \cite{vau00}. The author also thanks the Royal Society for the APEX award \texttt{APX\textbackslash R1\textbackslash 241139} which supported this research. 
\end{acknowledgments}


\appendix


\section{Derivation from a microscopic model}

\label{app:micro}

Here we derive our deme-level (mesoscopic) model (\ref{eqn:wf}) as the limiting form of a speaker-level (microscopic) copying and migration process. The goal is to make the connection between deme- and speaker-level behaviour explicit.  

\subsection{Microscopic model and master equation.}

Let $N_i$ be the number of sites in deme $i$.  We will assume that every site in the deme is occupied by a single speaker, and that each speaker uses one variant, making them either an $A$-speaker or an $a$-speaker . Let $Y_{ij}(t) \in \{0,1\}$ be the indicator that the speaker in the $j$th site of  deme $i$ at time $t$ is an $A$-speaker. We assume that the value of $Y_{ij}$ can \textit{update} due to `switching': the speaker changes variant, or `migration': the speaker leaves the deme and is replaced by a speaker from a different deme. We ignore within-deme residence swapping since it does not affect the variant counts within the deme. We model the birth and death of speakers by assuming that when a speaker dies they are replaced with a new speaker in the same site, who selects their state via a switching update. 

We define $M_i(t)$ to be the number of sites in deme $i$ which contain an $A$-speaker at time $t$
$$
M_i(t) = \sum_{j=1}^{N_i} Y_{ij}(t).
$$
We refer to $M_i(t)$ as the \textit{state} of the deme. Let $\bv{M}(t)=(M_1(t), \ldots, M_n(t))$ be the state vector all the demes. When defining switching dynamics, we distinguish between local updates which are influenced by other speakers in the deme, non-local updates which are influenced by other demes, and `mutations' where the speaker randomly switches variant.  During time $\dt$, we assume that each speaker will make a locally-influenced switch with probability $\lambda_1 \dt + o(\dt)$. Let $x$ be the fraction of $A$ speakers in a deme. We suppose that when making a locally-influenced update, speakers in this deme adopt variant $A$ with probability
\begin{equation}
g_1(x;s,\beta) = \frac{x}{x+(1-x)\exp\left(-s-\beta(1-2x)\right)}, 
\label{eqn:g1}
\end{equation}
where $s$ is the \textit{bias} parameter and $\beta$ is the \textit{conformity} parameter. The selection probability (\ref{eqn:g1}) has the symmetry  $g_1(x;s;\beta)= 1-g_1(1-x;-s;\beta)$. In the absence of bias and conformity we have $g_1(x;0,0)=x$, corresponding to voter dynamics on a fully connected contact network within the deme. When $\beta=0$, (\ref{eqn:g1}) reduces to the selection probability recently introduced in \cite{mon23} for inferring selection bias in the Wright-Fisher model of population genetics. When $s>0$ there is a selection bias in favour of variant $A$. When $s=0$ we obtain
$$
g_1(s;0,\beta)=\frac{x e^{\beta x}}{xe^{\beta x} + (1-x)e^{\beta (1-x)}}
$$
which biases selection in favour of the currently most popular variant.  We will assume that $s$ and $\beta$ are $O(N_i^{-1})$ so $g_1(x;s,\beta) = x + O(N_i^{-1})$. In this sense bias and conformity effects are $O(N_i^{-1})$ perturbations of within-deme voter dynamics. However, we will see below that their influence on system behaviour becomes increasingly strong as $N_i$ becomes large. 

We assume that for every speaker, the probability of a non-local update in time $\dt$ is $\lambda_2 \dt$. Writing the vector of $A$-speaker fractions in the demes as $\bv{x}=(x_1, \ldots, x_K)$, then when a speaker in deme $i$ performs a non-local update, they select variant $A$ with probability
$$
g_2(\bv{x};\bv{J}_i) = \sum_{j=1}^K J_{ij} x_j
$$
where $\bv{J}_i=(J_{i1}, \ldots, J_{iK})$, $J_{ij} \geq 0$, $J_{ii}=0$ and $\sum_{j=1}^K J_{ij}=1$. Here $J_{ij}$ represents the normalised influence of deme $j$ on deme $i$. We have the following symmetry $g_2(\bv{1}-\bv{x};\bv{J}_i) = 1- g_2(\bv{x},\bv{J}_i)$. We assume that $\lambda_2/\lambda_1 = O(N_i^{-1})$ so, like bias and conformity in local switching, non-local influences are an $O(N_i^{-1})$ perturbation of within-deme voter dynamics. Bias and conformity parameters are not included at the non-local level since, accordng to our assumptions, their contribution would be $O(N_i^{-2})$. The simplest switching update is mutation. During time $\dt$ we assume this occurs with probability $\nu \dt + o(\dt)$ for each speaker, where $\nu = O(N_i^{-1})$. A mutating speaker reselects their variable, choosing  $A$ or $a$ with equal probability.

Finally, we assume that each speaker leaves their deme with probability $\lambda_3 \dt$ in time $\dt$. When a speaker leaves deme $i$, let $m_{ij} \ge 0$ be the probability that their replacement comes from deme $j$, with $m_{ii}=0$, and $\sum_{j=1}^K m_{ij}=1$. The probability that the replacement will be an $A$-speaker is then $g_2(\bv{x},\bv{m}_i)$ where $\bv{m}_i=(m_{i1}, \ldots, m_{iK})$. According to these assumptions migration and non-local interactions are interchangeable in terms of their influence on the dynamics of the model. We cannot therefore separately infer interaction and migration parameters based on observations of deme states. We therefore discard migration terms from the model and reinterpret the non-local update rate and influence parameters as combining both non-local switching and migrations. In some situations migration flows may be estimated from census data, allowing us to separate the effects of influence from migration. 

According to our assumptions, the state of the speaker in site $j$ in deme $i$ obeys the following probabilistic update rule
\begin{align*}
&Y_{ij}(t+\dt) = \\
&\begin{cases}
    Y_{ij}(t)  &\tx{w.p. } 1-(\lambda_1+\lambda_2+\nu) \dt \\
    1  &\tx{w.p. }  \lambda_1 g_1 \dt +\lambda_2 g_2 \dt + \frac{\nu}{2} \dt \\
    0  &\tx{w.p. } \lambda_1(1-g_1) \dt + \lambda_2 (1-g_2) \dt + \frac{\nu}{2} \dt
\end{cases}    
\end{align*}
where we have used the abbreviations $g_1 = g_1(x_i;s,\beta)$ and $g_2 = g_2(\bv{x};\bv{J}_i)$. We define transition rates out of state $\bv{M}(t)=\bv{k}=(k_1,\ldots,k_K)$ into state $\bv{k} \pm \bv{e}_i$ where $\bv{e}_i$ is the standard unit basis vector in the $i$th coordinate direction,
$$
T_i^{\pm}(\bv{k}) = \lim_{\dt \ra 0} \frac{\PP\left(\bv{M}(t+\dt)=\bv{k}\pm \bv{e}_i\middle\vert \bv{M}(t)=\bv{k}\right)}{\dt}.
$$
To express these rates in simple form, we first define  $\Delta_i = 1/N_i$ and $\bv{\Delta}=(\Delta_1, \ldots, \Delta_K)$ and introduce the function
$$
\phi_i(\bv{k} \circ \bv{\Delta}) = g_1(k_i \Delta_i;s,\beta) + \frac{\lambda_2}{\lambda_1} g_2(\bv{k} \circ \bv{\Delta};\bv{J}_i) + \frac{\nu}{2 \lambda_1},
$$
where $\circ$ denotes the element-wise (Hadamard) product $\bv{k}\circ \bv{\Delta} = (k_1 \Delta_1, \ldots, k_K \Delta_K)$. We may then write transition rates in terms of this function as follows
\begin{align}
    T_i^+(\bv{k}) &= \lambda_1 (N_i-k_i) \phi_i(\bv{k}\circ \bv{\Delta}) \\
    T_i^-(\bv{k}) &= \lambda_1 k_i\left(1-\phi_i(\bv{k}\circ \bv{\Delta})+\frac{\lambda_2+\nu}{\lambda_1}\right).
\end{align}
The probability mass function over deme states is defined $p(\bv{k},t) = \PP(\bv{M}(t)=\bv{k})$, and satisfies the master equation
\begin{align}
\nonumber
\pa_t p(\bv{k},t)& = \\
\nonumber
\sum_{i=1}^K \Big( &T_i^+(\bv{k}-\bv{e}_i)p(\bv{k}-\bv{e}_i,t) + T_i^-(\bv{k}+\bv{e}_i)p(\bv{k}+\bv{e}_i,t) \\
& -(T_i^+(\bv{k}) + T_i^-(\bv{k}))p(\bv{k},t)\Big),    
\label{eqn:master}      
\end{align}
which arises from the law of total probability 
$$
p(\bv{k},t+\dt) = \sum_{\bv{k}'} \PP\left(\bv{M}(t+\dt)=\bv{k} \middle\vert \bv{M}(t)=\bv{k}'\right) p (\bv{k}',t).
$$
Our goal is now to show that the model (\ref{eqn:wf}) arises as a continuos state diffusion approximation to equation (\ref{eqn:master}).

\subsection{Diffusion approximation.}

We derive our diffusion approximation to (\ref{eqn:master}) via the Kramers-Moyal expansion \cite{gar09} in powers of $\Delta_i$. This is a standard technique, so we keep our derivation brief. We define the volume element $\dV = \prod_{i=1}^K \Delta_i$, and introduce a smooth probability density $f(\bv{x},t)$ defined for $\bv{x} \in [0,1]^K$, such that $ p(\bv{k},t) \approx  f(\bv{k} \circ \bv{\Delta})\dV $. We also define continuous state transition rate functions $W^{\pm} : [0,1]^K \ra [0,\infty)$, satisfying $W_i^{\pm}(\bv{k} \circ \bv{\Delta}) = T_i^\pm(\bv{k})$. We then approximate the first two terms in the summand on the right of the master equation (\ref{eqn:master}) using a Taylor series expansion of $W^{\pm}_i$ about the point $\bv{x} = \bv{k}\circ \bv{\Delta}$
\begin{align*}
    \frac{T_i^\pm(\bv{k}\mp \bv{e}_i)p_i(\bv{k} \mp \bv{e}_i,t)}{\dV}
    =& W_i^\pm(\bv{x} \mp \Delta_i \bv{e}_i) f(\bv{x} \pm \Delta_i \bv{e}_i) \\
    = W_i^\pm(\bv{x}) f(\bv{x},t) &\mp \Delta_i \pa_{x_i}(W_i^\pm(\bv{x}) f(\bv{x},t)) \\
    + \frac{\Delta_i^2}{2} \pa_{x_i}^2 &(W_i^\pm(\bv{x}) f(\bv{x},t)) + O(\Delta_i^3).
\end{align*}
Substituting into (\ref{eqn:master}), and keeping terms to order $\Delta_i^2$ we obtain the following Fokker-Planck equation
\begin{align*}
\pa_t f(\bv{x},t) =-&\sum_{i=1}^K \Delta_i \pa_{x_i} \left( (W_i^+(\bv{x}) - W_i^-(\bv{x})) f(\bv{x},t)\right)\\
&+ \sum_{i=1}^K \frac{\Delta_i^2}{2} \pa_{x_i}^2 \left( (W_i^+(\bv{x}) + W_i^-(\bv{x})) f(\bv{x},t)\right).    
\end{align*}
We now observe that 
\begin{align*}
\Delta_i (W_i^+(\bv{x}) - W_i^-(\bv{x})) = &\lambda_1\left(\phi_i(\bv{x})-x_i-\frac{\lambda_2+\nu}{\lambda_1} x_i \right) \\
\Delta^2_i (W_i^+(\bv{x}) + W_i^-(\bv{x})) = &\Delta_i \lambda_1 \Bigg(\phi_i(\bv{x})(1-2x_i) \\
&+ x_i + \frac{\lambda_2+\nu}{\lambda_1} x_i\Bigg).
\end{align*}
Since, by assumption, $g_1(x_i,s,\beta)$ is an order $\Delta_i$ perturbation of $x_i$, and $(\lambda_2+\nu)/\lambda_1 = O(\Delta_i)$ then the function
$$
\epsilon_i(\bv{x}) = \phi_i(\bv{x}) - x_i - \frac{\lambda_2+\nu}{\lambda_1} x_i
$$
is $O(\Delta_i)$. Written in terms of $\epsilon_i$, keeping terms to order $\Delta_i$ our Fokker-Planck equation is then
\begin{align*}
\pa_t f(\bv{x},t) =& - \lambda_1 \sum_{i=1}^K \pa_{x_i} (\epsilon_i(\bv{x}) f(\bv{x},t)) \\ 
&+ \sum_{i=1}^K \frac{\lambda_1}{N_i} \pa_{x_i}^2 (x_i(1-x_i) f(\bv{x},t)).    
\end{align*}
This equation describes the probability density of a family of coupled Wright-Fisher diffusion processes
$$
dX_i = \lambda_1 \epsilon_i(\bv{X},t) dt + \sqrt{\frac{2 \lambda_1 }{N_i}X_i(1-X_i)} dW_i
$$
where $W_1, \ldots, W_K$ are independent Brownian motions. The stochastic process $\bv{X}(t)$ has approximately the same distribution as the process $\bv{\Delta} \circ \bv{M}(t)$. Expanding $\epsilon_i(\bv{x})$ to linear order in the bias and conformity parameters, we obtain
\begin{align*}
\lambda_1 \epsilon_i(\bv{x}) =& \lambda_1 s x_i(1-x_i) + \lambda_1 \beta x_i(1-x_i)(2x_i-1) \\
&+ \frac{\nu}{2}(1-2x_i) + \lambda_2 \sum_{j=1}^K J_{ij} (x_j-x_i).    
\end{align*}
To simplify notation we redefine parameters $\lambda_1 s \ra s$, $\lambda_1 \beta \ra \beta$, $\lambda_2 J_{ij} \ra J_{ij}$, $\nu \ra 2 \nu$.

We have implicitly assumed that the network of within-deme interactions between speakers is \textit{fully connected}. In reality we expect speakers to interact only with a subset of their deme, and for different speakers to have different numbers of contacts, producing a \textit{heterogeneous} network. Analysis of voter models on heterogeneous networks \cite{ant06,soo07} reveals that the sizes of random fluctuations in the fraction of individuals in a given state is affected by network structure. In particular, greater heterogeneity leads to larger fluctuations. Since fluctuation sizes are determined by deme population $N_i$, we can account for network heterogeneity by replacing $N_i$ with an \textit{effective population size}, $N_{e,i}$ \cite{def21,wap25}, which is typically smaller than $N_i$. To keep our notation compact, we define the \textit{volatility} of deme $i$ in terms of the effective population, as $\sigma_i = \sqrt{2 \lambda_2 /N_{e,i}}$. To avoid over parameterisation, when fitting to data we will assume all demes have the same volatility.

Our final step is to promote the bias parameter to be a time dependent field, yielding the Wright fisher diffusion  (\ref{eqn:wf}). When $\beta=0$ this is formally identical to Kimura's stepping stone model \cite{Kim64,eth11} of genetic evolution with migration and space-time dependent selection. However, in our case the interaction coefficients measure both migration and influence.  


\section{Lifting matrix results}

\label{app:LA}

Here we provide brief proofs of results used in the derivation of the lifting matrix. We first note that the Gram matrix
$$
W_{ij} = \exp\left(-\frac{\Vert \bv{r}_i-\bv{r}_j\Vert^2}{2 \eta^2} \right)
$$
is symmetric and positive definite. This is a consequence of Schoenberg's theorem \cite{sch38}: If $f: [0,\infty) \ra \RR$ is completely monotonic, then the kernel defined by $K(\bv{r}_i,\bv{r}_j)=f(\Vert \bv{r}_i-\bv{r}_j\Vert)$ is a positive definite kernel. Thus, for any $\bv{y} \in \RR^n$, $\bv{y}^T \mat{W} \bv{y} >0$. Define the diagonal matrix $\mat{D}$ via $D_{ii}=\sum_{j=1}^n W_{ij}$ and the degree normalised Gram matrix $\bv{\Sigma} =\mat{D}^{-1/2} \mat{W} \mat{D}^{-1/2}$. Also define
$$
\bv{y} = \mat{D}^{-1/2} \bv{x}. 
$$
From the positive definiteness of $\mat{W}$ we have
$$
\bv{y}^T \mat{W} \bv{y} = \bv{x}^T \mat{D}^{-1/2} \mat{W} \mat{D}^{-1/2} \bv{x} = \bv{x}^T \mat{\Sigma} \bv{x} >0
$$
so $\mat{\Sigma}$ is also positive definite. By the triangle inequality $|a_1 + \ldots + a_n| < |a_1|+ \ldots + |a_n|$ and the fact that $|ab| \leq \tfrac{1}{2}(a^2 + b^2)$ we have
\begin{align*}
    \left\vert \bv{x}^T \mat{\Sigma} \bv{x} \right \vert &=  \left\vert \sum_{i,j} y_i W_{ij} y_j \right \vert \\
    &\leq \sum_{ij} \vert W_{ij} y_i y_j| \\
    & \leq \frac{1}{2} \sum_{ij} W_{ij}(y_i^2 + y_j^2) \\
    & = \sum D_{ii} y_i^2 = \bv{x}^T \bv{x}.
\end{align*}
Therefore the Rayleigh quotient $R(\mat{\Sigma},\bv{x}) = \bv{x}^T \mat{\Sigma} \bv{x}/\bv{x}^T \bv{x}$ satisfies
$$
0 < \left \vert R(\mat{\Sigma},\bv{x}) \right \vert \leq 1.
$$
Since the Rayleigh quotient (for a symmetric matrix) is bounded by the smallest and largest eigenvectors, with its maximum and minimum respectively achieving those values, then the eigenvalues of $\mat{\Sigma}$ must lie in the interval $(0,1]$.

\section{Interface shape and velocity}

\label{app:inter}

We first find a show that equation (\ref{eqn:interface}), given by
$$
D_{\tx{e}} x'' + \beta (2x-1)x(1-x) + \nu (1-2x) = 0
$$
admits a solution of the form $x(r_1)= \tfrac{1}{2} + a \tanh (br_1)$. Substitution into the above equation yields the condition
$$
-4 b^2 D_{\tx{e}} + \beta - 4 \nu - 
   4 (b^2 D_{\tx{e}} + a^2 \beta) \tanh^2(b r)=0.
$$
Since this must hold for any $r_1$ we have $-4 b^2 D_{\tx{e}} + \beta - 4 \nu=0$ and $b^2 D_{\tx{e}} + a^2 \beta=0$. Solving for $a$ and $b$ we have 
\begin{align*}
    a &= \frac{1}{2} \sqrt{1 - \frac{4 \nu}{\beta}}\\
    b &= \frac{1}{2}  \sqrt{\frac{\beta-4\nu}{D_{\tx{e}}}}.
\end{align*}
This solution --- a straight stationary interface aligned with the $r_2$ axis --- is an equilibrium solution to the partial differential equation (\ref{eqn:pde})
$$
\frac{\pa x}{\pa t} = \beta (2x-1)x(1-x) + \nu (1-2x) + D_{\tx{e}} \frac{\nabla^2 (\rho x)}{\rho}.
$$
Now let us consider interface solutions to this equation that are not in equilibrium. 
Consider a curved interface along which $x(\bv{r})=\tfrac{1}{2}$, having approximately the same profile as our straight interface. Let $P$ be a point on the interface and let $\hat{\bv{g}}$ be a unit vector which is normal to the interface at this point. Let $g$ be the magnitude of a displacement from $P$ along $\hat{\bv{g}}$. We have
\begin{align*}
    \nabla x &= \left(\frac{\pa x}{\pa g}\right)_t \hat{\bv{g}} \\
    \nabla^2 x &= \left( \frac{\pa ^2 x}{\pa g^2}\right)_t + \left(\frac{\pa x}{\pa g}\right)_t \nabla \cdot \hat{\bv{g}},
\end{align*}
where $(\bullet)_t$ denotes a derivative with $t$ held fixed. The diffusion term in our partial differential equation (pde) (\ref{eqn:pde}) may be expanded as
\begin{align*}
    \frac{\nabla^2 (\rho x)}{\rho} &= \nabla^2 x + x \frac{\nabla^2 \rho}{\rho} + 2 \frac{\nabla \rho \cdot \nabla x}{\rho} \\
    &= \left( \frac{\pa ^2 x}{\pa g^2}\right)_t + \left(\frac{\pa x}{\pa g}\right)_t \left( \nabla \cdot \hat{\bv{g}} + 2 \frac{\nabla \rho \cdot \hat{\bv{g}}}{\rho} \right) + x \frac{\nabla^2 \rho}{\rho}.
\end{align*}
We now make the assumption that the profile of the interface along the axis defined by $g \hat{\bv{g}}$ matches that of the equilibrium interface, so
$$
D_{\tx{e}} \left( \frac{\pa ^2 x}{\pa g^2}\right)_t + \beta (2x-1)x(1-x) + \nu (1-2x) = 0.
$$
The pde may then be written
$$
\left(\frac{\pa x}{\pa t}\right)_{g} = \left(\frac{\pa x}{\pa g}\right)_t D_{\tx{e}} \left( \nabla \cdot \hat{\bv{g}} + 2 \frac{\nabla \rho \cdot \hat{\bv{g}}}{\rho} \right) 
$$
where we made use of the assumption made when deriving the pde, that $D_{\tx{e}} |\nabla^2 \rho|/\rho \ll 1$. Finally, making use of the cyclic relation
$$
\left(\frac{\pa x}{\pa t}\right)_{g} \left(\frac{\pa g}{\pa x}\right)_t = - \left(\frac{\pa g}{\pa t}\right)_x
$$
and recognising $(\pa g/\pa t)_x$ as the velocity, $v$, of the interface, we have
$$
v = -D_{\tx{e}} \left( \nabla \cdot \hat{\bv{g}} + 2 \frac{\nabla \rho \cdot \hat{\bv{g}}}{\rho} \right). 
$$
This is relationship (\ref{eqn:v}) given in the main text.

\end{document}